\documentclass[nojss]{jss}\usepackage[]{graphicx}\usepackage[]{xcolor}
\makeatletter
\def\maxwidth{ %
  \ifdim\Gin@nat@width>\linewidth
    \linewidth
  \else
    \Gin@nat@width
  \fi
}
\makeatother

\usepackage{thumbpdf}
\usepackage{amsmath,amssymb}
\usepackage{amsfonts,amsthm}
\usepackage{amsmath}
\usepackage[round]{natbib}
\usepackage{bm}
\usepackage{rotating}

\newcommand{\R}{\mathbb{R}}

\theoremstyle{definition}

\newcommand{\nS}{q}

\newcommand{\class}[1]{`\code{#1}'}
\newcommand{\fct}[1]{\code{#1()}}

\extrafloats{100}

\setkeys{Gin}{width=\textwidth}

\title{Overlapped We Stand, Separated We Fall: Software Approaches to Diagnosing Separation Phenomena in Categorical Response Models}
\Shorttitle{Software Approaches to Diagnosing Separation in Categorical Response Models}

\author{Thomas Rusch\\WU Vienna University of Economics and Business
  \AND Lukas Sablica\\ WU Vienna University of Economics and Business
   \AND Kurt Hornik \\ WU Vienna University of Economics and Business}
\Plainauthor{Thomas Rusch, Lukas Sablica, Kurt Hornik}

\Abstract{In statistical models for categorical responses, including binary, nominal and ordinal responses, a phenomenon called ``separation'' may occur which can render maximum likelihood estimate nonexistent and/or nonunique. Based on \citet{Sablicaetal}'s general characterization of separation phenomena via structure vectors, in this article we discuss computational, numerical and practical aspects of separation phenomena in different software packages, including consequences, remedies and ways of diagnosing them. We further introduce the \proglang{R} package \pkg{divoRce} which provides the most comprehensive software suite to diagnose separation in existence. It allows to check for separation, to distinguish between types of separation, characterize the recession cone and identify variables and observations contributing to separation for almost any categorical response model in the literature, including models with baseline-category link, cumulative link, adjacent-category link and sequential link specifications. Diagnostics are based on methods for computational geometry and linear programming and can be done in floating-point or rational arithmetic with different solvers. The functionality is easily extendible by wrappers and S3 methods to accommodate implementations in third-party packages or of new categorical response models. We discuss our implementation in detail and show our software in action in a wide range of applications. We recommend to incorporate our checks for separation and existence of the MLE as a diagnostic routine in every categorical data analysis both in third-party software implementations and as part of a standard model diagnostic workflow.}


\Address{
  Thomas Rusch\\
  Competence Center for Empirical Research Methods\\
  WU (Vienna University of Economics and Business)\\
  Welthandelsplatz 1, Building D5\\
  1020 Vienna, Austria\\
  E-mail: \email{Thomas.Rusch@wu.ac.at}\\
}
\IfFileExists{upquote.sty}{\usepackage{upquote}}{}
\begin{document}
\section{Introduction}
When estimating parameters in models for categorical responses via
standard maximum likelihood estimation (MLE), it is possible to
encounter a phenomenon coined ``separation''
\citep[][]{polytomous:Albert+Anderson:1984}. Traditionally, the
literature distinguishes between two cases of separation, complete (or
perfect) and quasi-complete (near-perfect) separation, and the case of
overlap (no separation). Informally, separation occurs whenever there is
perfect separability (and thus prediction) of at least one response variable category for some linear combination of the explanatory variables. A simple example would be if for any value $x \geq x_0$ of a predictor (say, age in years) we only have observed a single category for a binary response $y$, say if $x \geq 40$ then $y=1$.

Separation can pose difficulties for model fitting and inference in a maximum likelihood setting. This is because when estimating model parameters from data that exhibit this property, the maximum likelihood estimate may effectively be infinite and therefore does not exist in the conventional sense. Additionally, standard errors as well as standard inferential procedures like Wald confidence intervals or $p$~values become nonsensical and numerical problems may occur. This can prove problematic if the separation phenomenon goes undetected and the parameter estimates and other inferential quantities are taken and interpreted at face value, or if the source of the estimation problem is misattributed.

We illustrate the situation with the data set \code{RecreationDemand} from \pkg{AER} \citep{AER} which is from a survey on whether people took a trip to Lake Somerville, Texas (encoded in the binary dependent variable ``TripYes''). As explanatory variables we have the categorical variable ``userfee'' (whether the person paid the annual user fee at Lake Somerville) and the numeric variables ``CostS'' (the cost of visiting Lake Somerville) and ``CostC'' (the cost of visiting Lake Conroe, Texas instead). We fit an additive logistic regression model (a binomial GLM with the logit link) 
\begin{Schunk}
\begin{Sinput}
> data("RecreationDemand", package = "AER")
> TripsBin <- RecreationDemand$trips > 0
> recdem <- cbind(RecreationDemand, TripYes = factor(TripsBin))
> m.recdem <- glm(TripYes ~ costS + userfee + costC, family = binomial,
+     data = recdem)
\end{Sinput}
\end{Schunk}
and look at the estimated coefficients, standard errors and $p$~values in the mode summary as well as the predicted probability for observation $530$ (a person who paid the annual user fee) at convergence of the model fitting algorithm.
\begin{Schunk}
\begin{Sinput}
> summary(m.recdem)
\end{Sinput}
\begin{Soutput}

Call:
glm(formula = TripYes ~ costS + userfee + costC, family = binomial, 
    data = recdem)

Coefficients:
             Estimate Std. Error z value Pr(>|z|)    
(Intercept)  -0.13537    0.15056  -0.899    0.369    
costS        -0.11446    0.01249  -9.164   <2e-16 ***
userfeeyes   15.98429  602.90823   0.027    0.979    
costC         0.11471    0.01245   9.212   <2e-16 ***
---
Signif. codes:  0 '***' 0.001 '**' 0.01 '*' 0.05 '.' 0.1 ' ' 1

(Dispersion parameter for binomial family taken to be 1)

    Null deviance: 866.53  on 658  degrees of freedom
Residual deviance: 724.21  on 655  degrees of freedom
AIC: 732.21

Number of Fisher Scoring iterations: 15
\end{Soutput}
\begin{Sinput}
> exp(coef(m.recdem)["userfeeyes"])
\end{Sinput}
\begin{Soutput}
userfeeyes 
   8747566 
\end{Soutput}
\begin{Sinput}
> p1logit <- predict(m.recdem, type = "response")[recdem$userfee ==
+     "yes"][1]
> p1logit
\end{Sinput}
\begin{Soutput}
      530 
0.9999999 
\end{Soutput}
\end{Schunk}
The profile confidence interval is given as
\begin{Schunk}
\begin{Sinput}
> confint(m.recdem)
\end{Sinput}
\begin{Soutput}
Waiting for profiling to be done...
\end{Soutput}
\begin{Soutput}
                  2.5 %      97.5 %
(Intercept)  -0.4294593  0.16250776
costS        -0.1397034 -0.09068746
userfeeyes  -25.5880695          NA
costC         0.0910208  0.13988101
\end{Soutput}
\end{Schunk}
In this data we have quasi-complete separation due to the category ``userfee = yes'': There was no-one who paid the annual user fee at Lake Somerville but did not visit it in that year. The consequence for estimation is that after a high number of $15$ iterations of the Fisher scoring algorithm, we get a coefficient of ``userfee'' of $15.98$ on the logit scale with a standard error of $602.91$, and an associated $p$~value of $0.98$ for the Wald test. The predicted probability of making the trip to Lake Somerville for observation $530$ is $0.9999999$. So while we have almost perfect prediction for some people in the data set (including observation 530) and an estimated $8.74 x 10^6$-fold higher chance of visiting Lake Somerville given one has paid the user fee (the odds ratio), the associated $p$~value is higher than any significance levels commonly employed. The profile confidence intervals returned may also be confusing because they output an \code{NA} value for the upper bound, and extend very far into the negative on the logit scale. This may lead analysts to falsely conclude that ``userfee'' is not an important variable or not significantly different from $0$, to remove the variable to ``fix'' the issues, or generally leave them in a confused state. This shows that if the problem of separation is not properly understood and diagnosed, it can present a pitfall for model fitting and, particularly, interpretation.

\paragraph{Related work.} So far in the literature, separation phenomena in categorical response models (CRM) have been discussed mainly from a formative viewpoint. In that understanding they are a property formed by a combination of individual aspects: the data which includes response and predictors, the model type which includes likelihood specification, link construction and link family specification (e.g., binary logistic regression model, proportional odds model, probit discrete choice model, etc.), and the specification of the linear predictor. Then, specific combinations are investigated. Examples of this formative view include \citet{polytomous:Albert+Anderson:1984,polytomous:Silvapulle:1981} who discuss the phenomena for binary logistic regression, \citet{polytomous:Kaufmann:1988} for some ordinal models with log-concave likelihoods, \citet{polytomous:Speckman+Lee+Sun:2009} for choice models with log-concave likelihoods, \citet{kosmidisACL} for adjacent-category logit models, \citet{schwendlogbinom} for logbinomial models. The formative view led to important results and insights for wide-spread use cases and to approaches for checking for separation in software, for example \citet{detectseparation} and \citet{schwendlogbinom}. Nevertheless, the formative view is theoretically limited.

Conversely, the structure vector framework of \cite{Sablicaetal} abstracts from the formative view by providing an integrative view based on a theory of ``structure vectors'' for the modle class of latent polyhedral categorical response models. Within this structure theory the mentioned individual aspects that comprise a CRM are condensed and encoded into a set of structure vectors that are unique for a given setup (similar to how a design matrix is unique for a given data situation). The theory then applies to any set of these structure vectors the same way, similar to how the theory on linear models applies to all design matrices the same way. This framework not only allows to characterize and diagnose overlap and separation for all existing categorical response model specifications based on a single criterion but also accommodates any future categorical response models as long as the corresponding structure vectors are found. Thus the structure vector theory of \cite{Sablicaetal} provides a general, integrative approach to separation in categorical response models. 

In this article, we focus on computational approaches to diagnosing separation in the wide class of categorical models derived from the multinomial likelihood family, including binary, ordinal and nominal categorical responses for almost all specifications and link functions available in the literature through the lens of the structure vector theory of \cite{Sablicaetal}. We relay the computational consequences of and software remedies for separation phenomena, discuss computational and numerical aspects of diagnosing separation phenomena, illustrate the approaches that are chosen by different software packages when encountering this phenomenon, and explicate how the structure vector theory can be turned into usable software. Based on the latter we will introduce the \proglang{R} package \pkg{divoRce} \citep{separationcheck} that offers a comprehensive diagnostic suite for almost all categorical response models in existence.

\section{Structure vector framework for categorical models}
We rely on \cite{Sablicaetal} who offer a comprehensive and unified treatment of MLE existence and uniqueness in models for categorical responses, that is for binary, categorical (aka nominal) and ordinal outcomes. Their theory covers binary, polytomous nominal and ordinal response cases for all link function constructions---baseline-category link (BCL), adjacent-category link (ACL), cumulative link (CL), sequential or continuation-ratio link (SL) models and fixed score ordered stereotype models (OS)\footnote{While the theory only works for fixed-score ordered stereotype models, \citet{Sablicaetal} establish a chain of sufficiency that allows to apply this also for the ordered stereotype models with estimated score parameters.}---and link family distributions (including the standard logit, probit, log, loglog, cloglog or cauchit links) that are commonly used. This unified and comprehensive treatment is achieved by characterizing separation or overlap in categorical response models as a geometric property of (a sequence of) structure vectors.

\subsection{Latent polyhedral categorical response models} 
As our statistical setup, we have $n$ independent $m$-dimensional random vectors $y_i = (y_{i1}, \ldots, y_{im})'$ which have multinomial distributions with probability vector $\pi_i = (\pi_{i1}, \ldots, \pi_{im})^\top$ and number $n_i = y_{i1} + \cdots + y_{im} \ge 1$ of underlying observations. It holds that $\pi_{ij} \geq 0$ for $j=1, \ldots, m$ and that $\sum_{j=1}^m \pi_{ij}=1$.

The $\pi_i$ depend on an unknown $p$-dimensional parameter vector $\beta$ with elements $\beta_b, b=1,\dots,p$ according to
\begin{equation}
  \label{eq:model}
  \pi_i = h\left(X_i \beta\right)=\left(h_1(X_i\beta), h_2(X_i\beta),\dots,h_m(X_i\beta)\right)^\top,
\end{equation}
with $X_i$ a known non-random design matrices of dimension $m \times p$ and $p$-dimensional row vectors $x_{i1}, \dots, x_{im}$. We note that this also accommodates linear basis expansions (e.g., for splines). The $h$  maps the relevant linear predictors into the probability simplex. When a baseline-category parameterization is used, one category may be assigned zero predictor. We further denote with $X$ the stacked $X_i$ for all $i$.

Following \citet{Sablicaetal}, let $[m]=\{1,\dots,m\}$ and $[m]_{-j}=[m]\setminus \{j\}$. For observation $i$, let $U_i\in\R^{r_i}$ have a distribution not depending on $\beta$. For each category $j\in[m]$, let $q_{ij}\geq1$, let $H_{ij}\in\R^{q_{ij}\times r_i}$ and let $B_{ij}=[b_{ij1},\ldots,b_{ijq_{ij}}]\in\R^{p\times q_{ij}}$ be known. A model is a latent polyhedral categorical response model if, for every admissible $\beta$ and up to latent events of probability zero, the sets
\begin{equation}
\mathcal{C}_{ij}(\beta)
=\left\{u\in\R^{r_i}:H_{ij}u\leq B_{ij}^{\top}\beta\right\}, \nonumber
\qquad j\in[m],
\end{equation}
form a partition of the latent space and
\begin{equation}
h_j(X_i\beta)
=\Prob\bigl(U_i\in\mathcal{C}_{ij}(\beta)\bigr). \nonumber
\end{equation}
The inequalities are coordinatewise. To connect this to the usual formulation of categorical response models, set $W_{ij}=H_{ij}U_i$ and let $G_{ij}$ be its CDF. Then the response probability can equivalently be written as
\begin{equation}
  h_j(X_i\beta)
  \label{eq:lpcrm}
=\Prob\bigl(W_{ij}\leq B_{ij}^{\top}\beta\bigr)
=G_{ij}\left(B_{ij}^{\top}\beta\right).
\end{equation}

This model class subsumes a large number of univariate and multivariate categorical response model classes for binary, categorical and ordinal outcomes including the most important and popular ones such as binary logistic and probit models, cumulative link models and baseline-category link models. See \cite{Sablicaetal}  for more details on how this translates to the model classes.

It is customary to estimate $\beta$ in such models with maximum likelihood. Using $I_j = \{ i : y_{ij} > 0 \}$ for the set of observations in category $j$ and neglecting terms not depending on $\beta$, the likelihood of $\beta$ is
\begin{equation}
    \label{eq:lik}
  L(\beta)
  =\prod_{i=1}^{n}\prod_{j=1}^{m}h_j(X_i \beta)^{y_{ij}}
  = \prod_{j=1}^{m}\prod_{i\in I_{j}} h_j(X_i \beta)^{y_{ij}}
\end{equation}
and the log-likelihood function takes the form
\begin{displaymath}
  \ell(\beta) = \sum_{j=1}^m \sum_{i \in I_j} y_{ij} \log(h_j(X_i\beta)).
\end{displaymath}

A maximum likelihood estimate of $\beta$, $\hat{\beta}$ (if it exists) is then
\begin{equation}
  \label{eq:ML}
 \hat{\beta}= \arg\max_{\beta \in \mathcal{B}} \ell(\beta)
\end{equation}
with $\mathcal{B}$ denoting the parameter space. Plugging in $\hat{\beta}$ into Equation~\ref{eq:model} gives the predicted probabilities $\hat{\pi}_{ij}$. Asymptotically, the variance-covariance matrix $\mathbb{V}(\hat{\beta})$ is given by
\begin{displaymath}
  \label{eq:vcovML}
 \mathbb{V}(\hat{\beta})\approx \mathbb{J}_n(\beta) = \mathbf{I}_n^{-1}(\beta) 
\end{displaymath}
with $\mathbf{I}_n(\beta)$ denoting the Fisher information matrix. The $b-$th diagonal element of $\mathbb{J}_n(\beta)$, $\mathbb{J}_n(\beta)(b,b)$ gives the estimated variance of $\beta_b$ and $\hat{se}_b = \sqrt{(\mathbb{J}_n(\beta)(b,b))}$ denotes the respective estimated standard error. 

It turns out that if the $\beta$ are estimated via Equation~\ref{eq:ML}, it can happen that the MLE does not exist (i.e., is infinite) at which point its standard error is undefined too due to a singular Fisher information matrix. One of the possible causes for this has been coined ``separation'' \citep{polytomous:Albert+Anderson:1984} and was already recognized by \citet{polytomous:Silvapulle:1981}.

\subsection{Structure vectors and separation}
\cite{Sablicaetal} characterize separation situations in latent polyhedral CRM rigorously, elegantly and general by showing that separation phenomena are closely linked to so-called structure vectors.

Structure vectors are vectors that encode how the faces of the latent events move with $\beta$ in Equation~\ref{eq:lik}. This includes the corresponding information available in the response $y_{ij}$, in the design matrix $X_i$, and the response function $h_j$ (which comprises both the link construction and the link family distribution).

We denote a structure vector as $s_i \in \R^p$, the non-empty finite set of structure vectors as $\mathcal{S} = \{ s_1, \ldots, s_{\nS} \}$ and the $\nS \times p$ matrix of structure vectors as $S$, with $s_i$ in the rows.

It turns out in latent polyhedral categorical response models, $s_i$ are the columns $b_{ijr}$ of the matrix $B_{ij}$ in Equation~{\ref{eq:lpcrm} and so
  \begin{equation}
    \mathcal{S}=\{b_{ijr}: j \in [m], i \in I_{j}, r \in [q_{ij}]\} \nonumber
\end{equation}
Then, we have can have three situations determined by whether and how the structure vectors can be separated by a hyperplane.

\paragraph{Overlap.} Overlap in the structure vectors occurs if for all nonzero $\beta\in \mathrm{span}(\mathcal{S})$, we have $s_i'\beta < 0$ for some $i \in [\nS]$. In other words the structure vectors cannot be separated by a hyperplane.

\paragraph{Complete separation.} We say that $\mathcal{S}$ has complete separation if there exists a (necessarily nonzero) $\beta\in \mathrm{span}(\mathcal{S})$ such that $
  s_i' \beta > 0, \quad \forall i \in [\nS]$. Here the structure vectors can be separated by a hyperplane with no vector lying exactly on the separating hyperplanes(s). 

\paragraph{Quasi-complete separation.} Finally, $\mathcal{S}$ exhibits quasi-complete separation if there exists a nonzero $\beta\in \mathrm{span}(\mathcal{S})$ satisfying $s_i'\beta \ge 0, \quad \forall i \in [\nS]$ with at least one equality. Compared to complete separation this is a weaker form of separation where some structure vectors may lie exactly on the separating hyperplane(s).

\cite{Sablicaetal} show that the structure vector definitions are equivalent to any number of prior definitions including the less general definitions of \cite{polytomous:Albert+Anderson:1984,polytomous:Silvapulle:1981,polytomous:Kaufmann:1988,polytomous:Speckman+Lee+Sun:2009}. The set of structure vectors not only contains all the information that is needed to check for separation phenomena but they also form the structure for the underlying geometry which can be used to obtain additional geometric insight into the separation situation, such as identifying the type of separation, characterizing the recession cone and its dimension, ascertaining which structure vectors or rows of $S$ are affected by separation and detecting which columns of $S$ are active in a separation direction. This can be done by utilizing computational geometry for calculations on the general convex polyhedron given by the structure vectors (details are in Sections~\ref{sec:linearities} and \ref{sec:rccone}) and by employing methods of linear programming for various linear programs for the structure vectors (details in Sections~\ref{sec:genlin} and \ref{sec:genlin2}). For the geometric calculations and linear programs to be accurate, especially for the case of quasi-complete separation, it may be worthwhile to employ a precision exceeding the standard double precision floating-point arithmetic. 

The following example illustrates structure vectors for a binary data situation with quasi-complete separation: For each $i$ we have a single response $y_{i} \in \{0,1\}$ and two scalar predictors $x_{1i}$ and $x_{2i}$, so $X_{i}=(1, x_{1i}, x_{2i})$ (the zero vector is omitted). A matrix of structure vectors for this setup is an $\nS=n \times p$ matrix with the rows being $s_i=-X_i$ if $i \in I_0$ and $s_i=X_i$ if $i \in I_1$, where the $X_i$ are the row vectors of stacked design matrix $X$. We can calculate this matrix $S$ in \proglang{R} by taking the rows of the design matrix \code{X} returned by \code{model.matrix} and multiplying all rows that correspond to $y_i=0$ with $-1$, 
\begin{Schunk}
\begin{Sinput}
> data("qcsepdat1")
> y <- qcsepdat1$y
> X <- model.matrix(y ~ x1 + x2, data = qcsepdat1)
> S <- X
> S[y == 0, ] <- -1 * S[y == 0, ]
> S
\end{Sinput}
\begin{Soutput}
   (Intercept)        x1           x2
1           -1 -1.000000  -3.00000000
2           -1 -2.000000   0.00000000
3           -1 -3.000000   1.00000000
4            1  3.000000   1.00000000
5           -1 -3.000000  -4.00000000
6            1  4.000000   0.00000000
7            1  5.000000   2.00000000
8            1  6.000000   7.00000000
9            1 10.000000   3.00000000
10           1 11.000000   4.00000000
11           1  7.923643  -0.71695576
12           1  5.841141   5.55543886
13           1  6.606504   0.80400287
14           1  9.176108   4.41846461
15           1  6.650658   5.92632068
16           1 10.345943  -0.60978838
17           1  9.516725   0.07964299
18           1 11.581141   4.66743639
19           1 10.809490  -0.24339052
20           1  8.031372   1.19162292
21          -1 -1.923957  -9.10082960
22          -1 -2.943028 -10.74461711
23          -1 -1.455745  -7.23534985
24          -1 -2.385796  -8.15634364
25          -1 -2.450655  -2.69138426
attr(,"assign")
[1] 0 1 2
\end{Soutput}
\end{Schunk}
Here w.l.o.g. we invert the vectors corresponding to category $0$ which can be adapted to any binary $y$. To illustrate, the structure vector corresponding to observation $5$ would be $(-1, -3, -4)^\top$. A couple of things are noteworthy for the binary situation: Firstly, $\nS=n$ in this case and $\dim(S)=\dim(X)$. This is not generally the case. Secondly, the $X_i$ of interest is a row vector of $X$ and corresponds to a single response $y_i$ in each row (thus $m=1$ as the zero vector reference is omitted). We will call this the ``compact long representation'' and this is how the design matrix typically looks like for \proglang{R} functions that fit categorical models, but not always. Thirdly, the above structure vectors would be appropriate for all link constructions, which is not generally the case.

A set of structure vectors can similarly be derived from the individual design matrix $X_i$ or stacked design matrix $X$ in combination with the responses $y_{ij}$ for other CRM and \cite{Sablicaetal} provide structure vectors for almost all CRM currently available in the literature, including the most widespread and popular ones.

We have implemented a dedicated generic function in \pkg{divoRce} to calculate structure vector matrices $S$ from a compact long representation for different CRM specifications. The function is \fct{structure\_vectors} and takes either a combination of a target vector \code{y} and a design matrix \code{X}, or a model formula with a data argument (see Section~\ref{sec:design} for details on the design and usage philosophy of our package) as well as an argument \code{model} referring to the link construction/model of interest, which currently is any of \code{"b"} (binary), \code{"bcl"} (baseline-category link), \code{"acl"} (adjacent-category link), \code{"os"} (ordered stereotype models) and \code{"cl"} (cumulative link). For example, for a cumulative link categorical response model \code{structure\_vector(y, X, model = "cl")} gives the structure vectors. Sequential models (continuation-ratio link, abbreviated as \code{"sl"}) can be expressed as a sequence of $S$ matrices for subsets of binary responses; for these \code{structure\_vectors} returns a list of length $m-1$ with each list element correponding sequentially to the categories $1, \dots, m-1$.

Table~\ref{tbl:models} gives an overview in terms of response type and the link construction of the categorical response model families that we cover and how we abbreviate them in this article and the software. Note that this applies to all link family distributions that meet the criteria of \cite{Sablicaetal}. The table also lists the \pkg{divoRce} functions that yield the structure vectors from a compact long representation.

\begin{table}
  \begin{tabular}{llll}
    \hline
    Response Type & Model Type/Link Construction & Abbreviation & Use  \\
    \hline
    Binary & all& b & \code{model = "b"} \\
    Categorical & baseline-category link& bcl & \code{model = "bcl"}\\
    Ordinal & cumulative link & cl & \code{model = "cl"}\\
                  & adjacent-category link& acl &  \code{model = "acl"}\\
                  & ordered stereotype models& os &   \code{model = "os"}\\
                  & sequential/continuation-ratio link& sl &  \code{model = "sl"}\\
     \hline
    \end{tabular}                                                           
  \caption{\label{tbl:models} The categorical response models that are currently supported in \pkg{divoRce}. Abbreviation refers to the string used to designate model class with the same link construction in both this article and the \pkg{divoRce} package. The last column, Use, features the argument that must be used to make \pkg{divoRce} functions work for the desired model class.}               
\end{table}

\section{Consequences and remedies}
\subsection{Consequences}
\label{sec:consequences}
If there is separation and thus the MLE is not finite, there can be a number of numerical and practical consequences that crop up when doing data analysis with standard software packages.  

First, standard implementations for fitting categorical response models will typically stop optimization at some point returning a finite value for some of the $p$ elements of $\hat{\beta}$, perhaps accompanied with a convergence warning. They are typically numerically large, but they need not be. Naturally, these values are all wrong---they should be $\hat{\beta}_b=\pm\infty$ for some $b \in \{1,\dots,p\}$. It may also happen that there is simply no convergence and model fitting procedure continues to run. 

Second, returned standard errors $\hat{se}_b$ can be hugely inflated. This is usually a numerical issue as theoretically the standard error is undefined but due to computing the standard error for a finite value at a point where the log-likelihood is almost flat, the numerical standard errors returned may be huge and numerically approach $\infty$. It may also be that no standard errors are returned and that one gets a warning that the Fisher information matrix/Hessian is singular. 
  
Third, from this it follows that standard frequentist inference based on the MLE may no longer be valid \citep{geyer2009likelihood}. For example, $p$~values for Wald test statistics will be near $1$ even though the effect size is infinitely large \citep[cf. Hauck-Donner effect,][]{hauckdonner}, and Wald confidence intervals will be extremely wide. Standard asymptotics for the likelihood ratio test statistic based on Wilk's theorem no longer applies, even though the maximized log-likelihood is numerically almost correct if the value of $\hat{\beta}_b$ that should be $\infty$ is numerically large \citep{geyer2009likelihood}. Confidence intervals based on the profile likelihood that invert the likelihood ratio or Score test may also be extremely wide, have numerical issues (e.g., for a the score version, the point estimate may not be in the CI), or are not computable.  

Fourth, there will be predicted probabilities of (almost) $1$ for some outcomes and some/all observations, including out-of-sample predictions. The associated certainty in prediction may be misplaced unless the separation phenomenon is also a characteristic of the data generating process (DGP). It should be noted here that separation is not a problem \emph{per se} in terms of the model or the data, as indeed perfect prediction is possible, but only a problem in terms of parameter estimation and inference. 

What further complicates matters is that while all of the above mentioned things can happen, not all of them need to happen in a given situation. We illustrate this with revisiting the example from Page 3 of the introduction, the trips to Lake Somerville. There we fitted a logistic regression model (the code is given in the introduction) and estimated the coefficient for ``userfee'' as $15.98$ with standard error of $602.91$, and $p$~value of $0.98$. The predicted probability of a ``1'' for ``TripYes'' for observation $530$ was $0.9999999$. For the logistic model, we see some indications for separation in the high value of the ``userfee'' coefficient and huge standard error. But at first glance the $p$~value looks legitimate (albeit given the effect size of the coefficient this should also be suspect) and \fct{glm} does not output a warning here that something might be off. At any rate, at least a few symptoms mentioned above are exhibited and separation may be inferred for this example. 

But things can change drastically when we also fit an additive relative risk regression model (i.e., a binomial GLM with the log link) to the data, see below. Some output is truncated here as there are $30$ warnings and we only display the last three that appear at the end of the iteration process (see Section~\ref{sec:Rdiag} for a more detailed discussion of the \fct{glm} warnings).
\begin{Schunk}
\begin{Sinput}
> m.recdemlog <- glm(TripYes ~ costS + userfee + costC, data = recdem,
+     start = c(-1, 0, 0, 0), family = binomial("log"))
\end{Sinput}
\begin{Soutput}
Warning: glm.fit: algorithm did not converge
\end{Soutput}
\begin{Soutput}
Warning: glm.fit: algorithm stopped at boundary value
\end{Soutput}
\begin{Soutput}
Warning: glm.fit: fitted probabilities numerically 0 or 1 occurred
\end{Soutput}
\end{Schunk}
\begin{Schunk}
\begin{Sinput}
> summary(m.recdemlog)
\end{Sinput}
\begin{Soutput}

Call:
glm(formula = TripYes ~ costS + userfee + costC, family = binomial("log"), 
    data = recdem, start = c(-1, 0, 0, 0))

Coefficients:
             Estimate Std. Error z value Pr(>|z|)    
(Intercept) -0.960427   0.052385 -18.334  < 2e-16 ***
costS       -0.012064   0.002497  -4.831 1.36e-06 ***
userfeeyes   0.721865   0.075851   9.517  < 2e-16 ***
costC        0.011383   0.002356   4.831 1.36e-06 ***
---
Signif. codes:  0 '***' 0.001 '**' 0.01 '*' 0.05 '.' 0.1 ' ' 1

(Dispersion parameter for binomial family taken to be 1)

    Null deviance: 866.53  on 658  degrees of freedom
Residual deviance: 809.29  on 655  degrees of freedom
AIC: 817.29

Number of Fisher Scoring iterations: 25
\end{Soutput}
\begin{Sinput}
> p1log <- predict(m.recdemlog, type = "response")[recdem$userfee ==
+     "yes"][1]
> p1log
\end{Sinput}
\begin{Soutput}
      530 
0.7989626 
\end{Soutput}
\end{Schunk}
Taking the output at face value, for this logbinomial model and after $25$ iterations of the IWLS algorithm we get a coefficient of ``userfee'' of $0.72$ with standard error of $0.08$, and $p$~value of $0$. The predicted probability of ``1'' for ``TripYes'' for observation $530$ is $0.8$. Here the returned coefficient and standard error of ``userfee'' look perfectly fine and the coefficient might be seen as significantly different from $0$ at the $5\%$ level. The predicted probability for observation $530$ also would not be suspicious, but since we have separation it is clearly too low given that perfect prediction is possible here. Still, none of these values need to raise any red flags and we could incorrectly conclude that the fit is unproblematic. The only indication that something went awry are the convergence problems: $25$ Fisher scoring iterations are a lot and there are various warnings during the fitting stage which are only available at runtime and not part of the object. The point of this comparison is that the mentioned consequences are neither specific nor sensitive enough to reliably diagnose whether a separation phenomenon is present or not. This is what we will discuss in Sections~\ref{sec:diag1} and \ref{sec:divorce}.

\subsection{Possible remedies}
If separation is encountered, there are different ways to deal with it:
\begin{enumerate}
\item If the analysis goal is prediction or if the MLE is finite, we may not do anything as (some of) our predictions are nearly perfect after all. In this case inference should be done cautiously and should not be based on the score function as e.g., the Wald test and corresponding CI may be subject to artefacts such as the Hauck-Donner effect. However, the predictor(s) associated with separation may then dominate other predictors. 

\item If the separation phenomenon encountered is not a property of the DGP but only due to the sample data (e.g., in case of rare events), then increasing the sample size can provide a fix as the chance of separation decreases with higher sample size (for a fixed number of predictors).

\item It is also possible to remove the observations that are subject to separation and fit the model for the subset of observations for which there is overlap. Clearly the data that were removed still need to be factored into the statistical and substantive interpretation afterwards, e.g., by a qualitative interpretation.

\item Another solution is for categorical predictors to collapse categories and for continuous predictors to bin values. Similarly, in case of a response with more than two categories, one can try collapsing response categories as well. Clearly, this must make sense in light of the data context and research question.   

\item There is a notorious suggestion that is sometimes made, and that is to get rid of the offending predictor variable(s) in the model. We do not recommended this as then highly predictive variables would be excluded from the model, knowingly creating a severe omitted variable bias.

\item There also exist statistical solutions to tame the numeric problem of infinite MLE. The main remedy is to use regularization/penalization/shrinkage estimation to keep the estimate from becoming infinite. A popular approach is to use Firth's correction which places a small penalty on the MLE \citep{firth1993bias,heinze2002solution,kosmidis2021}. This can be done in \proglang{R} with package \pkg{logistf} \citep{logistf} or \pkg{brglm2} \citep{brglm2}, in \proglang{SAS} with adding the \code{FIRTH} option in the call to \code{PROC LOGISTIC} and in \proglang{Stata} with \pkg{firthlogit} \citep{firthlogit}. This is equivalent to fitting a model with the Jeffreys prior. Other forms of regularization in a frequentist setting are to use lasso, ridge or elastic net penalties \citep{zou2005regularization}. It is also possible to use a Bayesian approach which introduces regularization via the prior. Specific Bayesian suggestions for separation are using weakly informative priors, for example Cauchy \citep{Gelman2008}, log-F(1,1) \citep{Greenland2015} or the half-prior \citep{Rainey2016}. If data sizes are small, using exact logistic regression will also help, available for example in \proglang{R} package \pkg{elrm} \citep{elrm}, in \proglang{SAS} with adding the \code{EXACT} option in the call to \code{PROC LOGISTIC}) and in \proglang{Stata} with \pkg{exlogistic}. We point out that there is the view that these remedies do not really solve the inference problems \citep{geyer2009likelihood}. 
\end{enumerate}
  
If a data analyst is familiar with separation phenomena and its effects, it may be possible to infer what is going on from the parameter fitting process, the model output and possible warnings and take the necessary steps to mitigate the problem. But as we saw in Section~\ref{sec:consequences}, diagnosing separation that way is not very reliable even for experienced data analysts. Furthermore due to the widespread use of categorical response models, it is possible that they are used by practitioners that are not aware of separation phenomena and may just proceed with interpreting coefficients, $p$~values and CIs or the predicted probabilities at face value.

Thus as a necessary prerequisite to be able to take any of the mentioned steps to mitigate the separation problem, one needs to be able to properly and reliably diagnose whether separation has been encountered and to identify which parts of the model (predictors, observations, link specification) are responsible. In the next section, we will discuss what opportunities and mechanisms for diagnosing separation have existed in third party software packages up to now. In Section~\ref{sec:divorce}, we describe our own \proglang{R} package \pkg{divoRce} for separation diagnostics.   

\section{Diagnosing separation in third-party software}
\label{sec:diag1}
The problem of separation has been recognized in many stastical software packages as something to diagnose and warn the user about. The way this is done however, and how well this works, differs considerably between software packages (and sometimes even between different subroutines of the same software). In the following we will discuss how this looks like on the user level for the most popular third-party software packages (\proglang{SAS}, \proglang{R}, \proglang{SPSS}, \proglang{Stata}), give algorithmic details and document the information available to us. An overview of the capabilities of different software packages with respect to separation phenomena is given in Table~\ref{tbl:software}; it includes \proglang{R} package \pkg{detectseparation} \citep{detectseparation} and the \pkg{divoRce} \proglang{R} package described in Section~\ref{sec:divorce}.  
\begin{sidewaystable}
  \begin{tabular}{lllllll}
    \hline
           &          & \proglang{SAS}/\proglang{SPSS}& \proglang{Stata}    & \proglang{R}:\fct{glm}       & \proglang{R}:\pkg{detectseparation}& \proglang{R}:\pkg{divoRce} \\
\hline
    Check  &                                     &         &          &             &               &       \\
           & baseline-category link (binary)     & yes     &yes       & partial     & yes           &yes   \\
           & baseline-category link (polytomous) & yes     &yes       & partial     & partial       &yes \\
           & cumulative link                     & yes     &yes       & no          & no            &yes \\  
           & adjacent-category link                  & partial &yes       & limited     & partial       &yes \\
           & stereotype model                    & partial &yes       & limited     & limited       &yes \\ 
           & sequential link                     & partial &yes       & limited     & limited       &yes \\
 \hline   
   Usage   &          &         &          &             &               &      \\
           & pre-fit  & yes     &yes       & yes         &yes            &yes    \\
           & post-fit & no      &no        & no          &yes            &yes   \\
 \hline   
  Features &                     &         &         &       &               &        \\
           & case distinction    & partial & partial & no    & yes            &yes \\  
           & row detection       & partial  &partial & no    & no            &yes \\ 
           & column detection    & partial &partial  & no    & yes           &yes \\ 
           & recession cone      & no      & no      &no     & no            &yes \\ 
           & rational arithmetic & no      & no      &no     & no            &yes \\
           & different solvers   & no      & no      & no    & yes            &yes \\
           & extensibility       & no      & limited & limited & yes     &yes \\
           & free open source    & no      & no      & yes    & yes          &yes \\
    \hline
  \end{tabular}
  \caption{\label{tbl:software} An overview of software capabilities for checking separation for \proglang{SPSS}, \proglang{Stata}, \proglang{SAS} and \proglang{R}. Here ``partial'' means not fully supported and ``limited'' means one needs to make much effort. For example for the separation check, ``partial'' means the check is only available with a specific model specification, e.g., fitting a multinomial logistic model via the Poisson trick or using \code{PROC NLMIXED} with manual link functions and ``limited'' means that it is available after data restructuring and specific model specification or code adaptation (e.g., splitting data and fitting a series of logistic regressions for sequential models).}  
\end{sidewaystable}

It is useful to classify the checks into pre-fit (where a model has not yet been fitted or is in the process of being fitted so may also refer to checks at runtime of the model fitting function) and post-fit (where a model has already been fitted and/or a model object is already available). 


\subsection{SAS}
\proglang{SAS} uses a pre-fit approach based on predicted probabilities using floating-point arithmetic. It is available in the \code{PROC LOGISTIC} \citep{sas-sep} and works for binary, multinomial and cumulative link models. It is not very reliable to detect quasi-complete separation \citep{allison}.

The approach is \citep{allison}: 
\begin{itemize}
  \item If the convergence criterion is satisfied within eight iterations overlap is concluded and no specific message is returned.  
 \item For all iterations after the $8$-th the predicted probabilities $\hat{\pi}_{im}$ for all categories and all observations are calculated. If for all $i$ there is a $\hat{\pi}_{ij} \geq 0.999$ for any $j=1,\dots,m$, then complete separation is inferred and a warning is printed.
 \item If for some $i$ the $\hat{\pi}_{ij} \geq 0.95$ for any $j$, the standard errors are examined. If they exceed some criterion (the exact one we could not find documented) then quasi-complete separation is inferred and a warning is given. 
\end{itemize}

A typical warning as printed in \proglang{SAS} can be found in Figure~\ref{fig:saswarn} (for quasi-complete separation). 
\begin{figure}
\centering  
\includegraphics[width=0.8\textwidth]{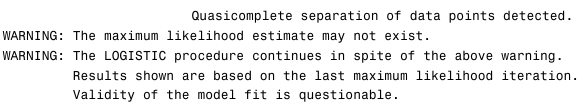}
\caption{\label{fig:saswarn} A typical warning for separation in \proglang{SAS}.}
\end{figure}

For other ordinal models in \proglang{SAS}, one can use \pkg{PROC NLMIXED} to fit adjacent-category, ordered stereotype and sequential models. \proglang{PROC NLMIXED} does not seem to have a dedicated separation check.

\subsection{SPSS}
\proglang{SPSS} uses a pre-fit approach based on predicted probabilities using floating-point arithmetic that is similar to the approach taken in \proglang{SAS}. It is available in the modules for choice models, binary, ordinal and multinomial regression via Genlin \citep{spss24}. 

The approach is roughly \citep{spss24}: 
\begin{itemize}
\item Starting with iteration \code{CHKSEP}, for each $i$ the $\pi^{max}_i=\max_j \hat{\pi}_{ij}$ is found.  
\item If $\pi^{max}_i >0.99$ it is assumed there is a perfect prediction for observation $i$.   
\item If all $i$ have perfect prediction it is concluded that there is complete separation and a warning is returned.
\item If some $i$ have perfect prediction and the Hessian of $\hat{\beta}$ at iteration \code{CHKSEP} is singular, it is concluded that there is quasi-complete separation and a warning is returned.
\item In every other case, overlap is seen as existing for the data. 
\end{itemize}

A typical warning as printed in \proglang{SPSS} can be found in Figure~\ref{fig:spsswarn} (for quasi-complete separation). 
\begin{figure}
\centering  
\includegraphics[width=0.9\textwidth]{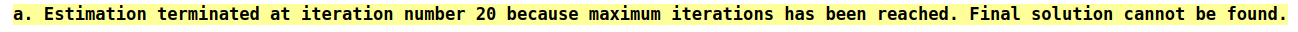}
\caption{\label{fig:spsswarn} A typical warning for separation in SPSS.}
\end{figure}

\subsection{Stata}
\proglang{Stata} has a built-in routine that checks for separation pre-fit. This is done internally via a call to a function \code{_rmcoll} whose source code is not available and the exact implementation is not documented\footnote{We asked the online community ``Statalist'' about it and obtained the same information at the time of writing. See \url{https://www.statalist.org/forums/forum/general-stata-discussion/general/1682249-how-does-stata-check-for-quasi-complete-separation?view=thread}}. 

At any rate, \proglang{Stata} prints a warning if it found separation. For quasi-complete separation it mentions which predictor combination is responsible for the almost perfect prediction and then by default automatically removes both the columns in the design matrix that are active in a separation direction as well as the observations associated with separation, and then proceeds with the estimation on the subset of the data. It does that for both categorical and continuous predictors.

A typical warning as printed in \proglang{Stata} can be found in Figure~\ref{fig:statawarn} (for quasi-complete separation). 
\begin{figure}
\centering  
\includegraphics[width=0.6\textwidth]{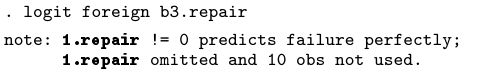}
\caption{\label{fig:statawarn} A typical warning for separation in \proglang{Stata}.}
\end{figure}
In case of complete separation it may also be that \proglang{Stata} warns \code{foo predicts outcome perfectly} and stops because, apparently, ``statistics dictates that no model can be fit.'' \citep[][p. 6]{rlogit}

\subsection{R}
\label{sec:Rdiag}
In \proglang{R} there are currently two ways of checking for separation, one that is built into \fct{glm} and one that can be found in the contributed package \pkg{detectseparation}. The latter provides checks for binary binomial GLM with different link function and for categorical response models that can be fit with \fct{glm}, including baseline-category logit and adjacent-category logit models. 

In the \pkg{stats} package which is part of base \proglang{R}, the model fitting function \fct{glm} includes a check based on predicted probabilities and them being numerically $0$ or $1$. It may give any or both of the two warning(s) at the end of the iteration process,
\begin{Schunk}
\begin{Soutput}
Warning messages: 
 1: glm.fit: algorithm did not converge 
 2: glm.fit: fitted probabilities numerically 0 or 1 occurred
\end{Soutput}
\end{Schunk}
with the latter being a bit more specific for a separation problem. Additionally \fct{glm} may return warnings related to problems during the iteration process like truncation of the step size due to divergence. Note that these warnings are only printed during runtime of \fct{glm} and there is no specific warning when a fitted object is printed or summarized.  

The \pkg{detectseparation} \citep{detectseparation} package implements thorough checks for separation. To date these are the most comprehensive checks available in \proglang{R} (and also beyond). 
The main functions are:  
\begin{description}
\item \fct{check\_infinite\_estimates}: The model is successively refitted with an increasing maximum number of allowed iterations. At each step, the ratio of the estimated standard errors at this maximum number of iterations and the standard error of the first fit is taken. If the sequence of ratios diverges the function concludes that the MLE is infinite. Thus this is a post-fit check. The function is an S3 generic function, with an S3 method for objects returned by \fct{glm} and \fct{brmultinom} or \fct{bracl} in \pkg{brglm2} (in case of the latter two this is restriction to logit links). 
\item \fct{detect\_separation}/\fct{detect\_infinite\_estimates}: Solves a linear program to check for any type of separation with different solvers and identifies which parameters have infinite maximum likelihood estimates. This is a pre-fit method and works either by supplying a design matrix and corresponding response vector, or via \fct{glm}. For the latter it can be activated in the call to \fct{glm} by using the \code{method} argument that then overloads the standard fitting function \code{glm.fit}. It also allows to to distinguish between the different types of separation for diagnostics with setting \code{separation_type = TRUE}.
\end{description}
Particularly with the latter function, this package features the most reliable and best checks available in any software up to this point. 

There are some disadvantages at the time of writing though: First, the functionality currently only extends to binary models (\class{glm} with a binomial family) and to models that can be expressed as a poisson GLM (e.g., baseline-category logit and the adjacent-category logit model) but not to general polytomous or ordinal responses with arbitrary link functions. 
 Additionally, there are currently no methods for \fct{check\_infinite\_estimates} for more classes. The package authors are working on that. Second, the linear program for \fct{detect\_separation} is currently implemented to happen mainly before model fitting but one can check an existing \class{glm} object post-fit via \code{update(object, method = "detect_separation")} with the linear program or via the heuristic implemented in \fct{check\_infinite\_estimates}. Third, little extra information about the situation is returned and fifth, numerical accuracy is restricted to floating-point arithmetic. In the next section, we present our package \pkg{divoRce} that complements the functions in \pkg{detectseparation} by remedying the disadvantages of the latter.

\section{divoRce: A comprehensive diagnostic suite for R}
\label{sec:divorce}
Our goal was to develop a general purpose diagnostic suite for separation phenomena in latent polyhedral categorical response models. This culminated in the \proglang{R} package \pkg{divoRce}. It is meant to complement and subsume the separation related functionality available in \proglang{R} to this date and enhance it to pre- and post-fit. We further enhance existing check functionality available in \proglang{R} to all ordinal model classes and almost all nominal model classes for any link construction, that is, baseline-category link, adjacent-category link, cumulative link, sequential or continuation-ratio link and ordered stereotype model. This further works for any link family distribution setup covered in \cite{Sablicaetal}, which includes any link derived from a continuous distribution function such as logit, probit, cauchit, loglog, cloglog or log link. Moreover, any hitherto nonexisting models can be diagnosed with \pkg{divoRce} functionality as long as structure vectors can be derived and constructed. Thus we not only provide a software suite for checking for separation but also a computational scaffolding that can be used to write wrappers and methods for implementations of categorical response models in third-party packages, or even for new categorical response models. Note that for link family distributions with bounded support, separation does not need to mean non-existence.

Different from the approach chosen in \pkg{detectseparation}, we decided to not make any changes to or overload functions in \pkg{base}, \pkg{stats} or contributed packages. We instead provide \pkg{divoRce} as a stand-alone contributed package that features an S3 object-oriented system that is extensible with methods and wrappers. Our package features support for already fitted models (post-fit) and does not rely on any model fitting functionality if pre-fit functions are used. 

\subsection{Software design}
\label{sec:design}
We implemented functions for pre-fit diagnostics as well as post-fit diagnostics 
in a user-friendly way via methods for a number of S3 generics. These generic functions are the high-level functions that we recommend to be primarily used. They are called \fct{check\_separation}, \fct{check\_overlap}, \fct{separation\_columns}, \fct{separation\_rows}, \fct{diagnose\_separation}, and finally \fct{recession\_cone}. 
Table~\ref{tbl:functions} provides an overview of the functions available.

\begin{table}
  \begin{tabular}{lll}
    \hline
    Description & Function Call & Usage \\
    \hline
    Separation check via \ref{lp1} & \code{check\_separation()} & pre/post-fit\\
    Overlap check via \ref{lp1} & \code{check\_overlap()} & pre/post-fit \\
    Separation check via \ref{lp2}      & \code{check\_separation(..., quick = TRUE)} & pre/post-fit\\
    Overlap check via \ref{lp2}      & \code{check\_overlap(..., quick = TRUE)} & pre/post-fit \\
    Sequential overlap check           & \code{check\_overlap(..., sequential = TRUE)} & pre/post-fit\\
    Parallel overlap check           & \code{check\_overlap(..., parallel = TRUE)} & pre/post-fit\\
    Detect separation columns & \fct{separation\_columns} & pre/post-fit \\
    Detect separation rows & \fct{separation\_rows} & pre/post-fit\\
    Characterize recession cone & \fct{recession\_cone} & pre/post-fit\\
    Identify linearities & \fct{linearities} & pre-fit \\
    Detailed diagnostics & \fct{diagnose\_separation} &  pre/post-fit \\
    \hline
    \end{tabular}                                                           
  \caption{\label{tbl:functions} Overview of the functions available in \pkg{divoRce}. The extra arguments give the specification to toggle the desired behavior. They can also be combined, e.g., \code{check\_overlap(..., quick = TRUE, parallel = TRUE)}. The third column designates whether the function can be used pre-fit, post-fit or both.}        
\end{table}

The functions can be used to answer the following questions:
\begin{itemize}
\item Whether there is separation or not: This functionality is found in the function \fct{check\_separation} which includes the standard check via the linear program in Equation~\ref{lp1} and quick check via the linear program in Equation~\ref{lp2}, 
  the function \fct{check\_overlap} featuring an overlap check via the standard check via the linear program Equation~\ref{lp1} and quick check via linear program in Equation~\ref{lp2} and possibly done sequentially or in a parallel with multiple cores, and in \fct{diagnose\_separation} that gives detailed diagnostics of the situation. 
\item What type of separation we have: \fct{diagnose\_separation} for detailed diagnostics.
 \item Which columns in the design matrix/structure vector matrix are active in a separation direction: Here we have the function \fct{separation\_columns} gives the names as well as the column indices column of the design matrix/structure vector matrix.
 \item Which rows in the design matrix/structure vector matrix are responsible for separation: The function \fct{separation\_rows} allows to get the data frame/submatrix as well as the row index of the rows of the design matrix/structure vector matrix and the outcome associated with separation. \fct{linearities} returns the set of structure vectors that show overlap in the negative structure vector matrix. The function \fct{diagnose\_separation} also returns the separation rows as part of the returned object in \code{$offrows}.
\item How the recession cone looks like: The generic \fct{recession\_cone} for the recession cone and its dimensionality. The results of \fct{diagnose\_separation} includes the recession cone dimension as part of the returned object \code{$offcols} as well.
\item Identification of the linearities:  \fct{linearities} returns the set of structure vectors in the negative structure vector matrix that are linearities (i.e., associated with overlap). 
\end{itemize}
We will describe the functions in detail in Sections~\ref{sec:genlin}-\ref{sec:fulldiag}. Depending on the setting, the functions in \pkg{divoRce} can be used in different ways.

\paragraph{Pre-fit usage.}
Pre-fit diagnostics means that no model has been fitted yet. In a pre-fit setting the checks and diagnostics are conducted either (a) for a combination of response vector (argument \code{y}), stacked design matrix (argument \code{X}) and model (specified as a character string via the argument \code{model}, one of \code{c("b", "bcl", "acl", "sl", "os")}, or (b) for a matrix of structure vectors (argument \code{S}), or (c) for a formula.

In case of (a), the \code{y} argument is used to dispatch on to a method for the class of \code{y} and the structure vector matrix is built internally from the three arguments \code{y}, \code{X} (which do not have a default value) and \code{model}. We expect the \code{X} to be a standard design matrix in the compact long representation as created in \proglang{R} with \fct{model.matrix} or \fct{model.frame}. If the \code{model} argument is missing we default to a standard model family for the type of response: \code{"b"} if the outcome is binary (numeric, integer, character, boolean or factor with $2$ distinct values or levels;), \code{"bcl"} for categorical (numeric, integer, character or unordered factor with $>2$ distinct values or levels) or \code{"cl"} for ordinal (ordered factor with $>2$ levels). We recommend to already supply \code{y} as an (ordered) factor to make sure that the reference/baseline level is registered correctly for the given \code{X}; else we internally convert \code{y} to factor with the lowest alpha-numeric value as the reference level. In \pkg{divoRce} we generally follow the convention in \proglang{R} to use the lowest value as reference.

In case of (b) a matrix argument to \code{S} leads to dispatch to a method for class \code{matrix}. Here the matrix \code{S} is the only necessary argument and is taken at face value. If an \code{S} argument is supplied, any argument to \code{y}, \code{X}, or \code{model} is ignored. In this case it falls within the responsibility of the user to provide the correct structure vector matrix.

For (c) we also offer methods for an object of class \class{formula} that takes the same formula/data combination as a fitting function would, plus the model argument as described above and then internally rebuilds to use variant (a). The formula is processed in the same way as in \fct{stats::glm}.   



We believe for most users using the generics via path (a), or via the formula interface (c) will be most useful. However, path (b) and the object-oriented mechanism easily allows third-party package authors to provide wrappers and methods for their own implementations of categorical response models. See below for an example.
 
\paragraph{Post-fit usage.} For post-fit diagnostics, a model has been already been fitted and an \proglang{R} model object \code{modelobject} is available. The primary use case is when a fitted model looks off somehow and the analyst suspects it may be due to separation, or if they just generally want to check the fitted model \emph{post hoc}.

Post-fit diagnostics are implemented via the S3 generic function mechanism described above, and the check is conducted via \code{foo(modelobject, rational)}; dispatch is based on \code{class(modelobject)} to an appropriate method. Currently implemented are post-fit check methods for the classes returned by \fct{stats::glm}, \fct{nnet::multinom}, \fct{MASS::polr}, \fct{ordinal::clm}, \fct{brglm2::bracl}, \fct{brglm2::brmultinom} and \fct{clustord::osm}.

Methods for classes not yet supported (or wrappers more generally) can easily be written by package authors by wrapping the generic pre-fit functions, most conveniently by utilizing the \code{S} argument. For such a wrapper, the $S$ matrix of structure vectors that is appropriate for the polyhedral latent categorical response model needs to be calculated in the code body, typically by calling a function to that effect. This result can then be passed as the \code{S} argument to any of the generic functions to dispatch to the \class{matrix} method. To illustrate this, we show how one can provide a \fct{check\_separation} method for the class \code{osm} of package \pkg{clustord} which allows to fit ordered stereotype models. First, we need to provide a function to generate and return the appropriate $S$, which we call \code{osm\_S}. In this case (and for all other cases listed in Table~\ref{tbl:models}) this is straightforward given that we already have a prepared function in \pkg{divoRce} that returns $S$ for OSM models; for didactic reasons we just wrap that one to a new function \code{osm_S}. For models not listed in Table~\ref{tbl:models} writing a function to obtain $S$ is necessary. 
\begin{Schunk}
\begin{Sinput}
> osm_S <- function(y, X, rational = FALSE) {
+     structure_vectors(y, X = X, rational = rational, model = "os")
+ }
\end{Sinput}
\end{Schunk}
With a function to get the correct $S$ matrix, we can utilize \fct{check\_separation} via the \code{S} argument. In the body of the \class{osm} method we need to extract the right parts of the \class{osm} object, reconstruct them and build the proper arguments to \fct{osm\_S}. The resulting \code{S} we then pass to the generic function \fct{check\_separation} which will dispatch to the \class{matrix} method.   
\begin{Schunk}
\begin{Sinput}
> check_separation.osm <- function(osmobject, rational = FALSE,
+     quick = FALSE, backend = c("rcdd", "ROI"), solver = NULL,
+     ...) {
+     y <- model.frame(osmobject)[, 1]
+     X <- model.matrix(osmobject)
+     S <- osm_S(y = y, X = X, rational = rational)
+     check_separation(S = S, rational = rational, quick = quick,
+         backend = backend, solver = solver, ...)
+ }
\end{Sinput}
\end{Schunk}
This can then be repeated for every generic function.

\paragraph{Rational arithmetic.}
\label{sec:rational}
One distinguishing feature of our software is that all the calculations can be made numerically as accurate as possible because \pkg{divoRce} functions can employ infinite precision rational arithmetic as offered in the package \pkg{rcdd} \citep{rcdd}. This is particularly useful for quasi-complete separation where we need to check for exact $0$. All functions in \pkg{divoRce} allow to use rational arithmetic with the flag \code{rational = TRUE}.

The package \pkg{rcdd} is the \proglang{R} interface to \pkg{cddlib} \citep{cddlib} which can use the GNU multiple precision library \citep[GMP;][]{GMP}. To make use of this feature in \pkg{rcdd} it suffices for numbers to be represented as rationals via specifying them as character strings \citep{geyervinny}. For example if one wants to represent the numeric \code{-0.5} as rational that way, it must be specified as the character \code{"-1/2"}. If all numbers that are input to functions from \pkg{rcdd} are represented that way, calculations can be done using arbitrary precision. This carries over to all the functions in \pkg{divoRce}, which allow for three situations:
\begin{itemize}
\item Floating-point arithmetic on floating-point input (default): Can be toggled via the argument \code{rational = FALSE}. It takes numbers represented in single precision floating-point format (the \proglang{R} standard) and does the calculations in floating-point arithmetic.
\item Rational arithmetic on floating-point inputs: Can be toggled via the argument \code{rational = TRUE}. It takes numbers represented in floating-point format and converts them internally to the rational format via \fct{rcdd::d2q}. Then it does the calculations in rational arithmetic. Numeric output is again in floating-point format (converted via \fct{rcdd:q2d}). 
\item Rational arithmetic on rational inputs: Is automatically toggled if the data are given in rational format and can be also manually toggled via the argument \code{rational = TRUE}. It takes numbers represented in rational format and does the calculations in rational arithmetic. Results from calculations will be returned in the rational representation.
\end{itemize}

Typically the data will be available in standard \proglang{R} data structures such as data frames or matrices with any of floating-point numbers, integers, characters, logical or factors. Model fitting functions either expect a design matrix $X$ and outcome $y$, or create them via formula from the data in floating-point representation; we do not know any that will work with the rational representation. Because of that, all our implemented functions can operate both on an outcome vector and design matrix pair $(y,X)$, or on the structure vector matrix $S$ for a given model for a given data set. This allows the user to supply them either in floating-point representation or rational representation.

From a computational point of view it is best for a user who wants to make full use of rational arithmetic to set up the $X$, $y$ or $S$ already as character in rational format and not rely on built-in conversion. This is due to the conversion still carrying over the limited precision. For example, a floating point value of \code{0.1} is supposed to be \code{"1/10"} but floating-point numbers of the decimal system are inexact due to computers using the binary system and the value of \code{"1/10"} cannot naturally be represented as a finite sum of powers of $2$ \citep{geyervinny}. The rational representation of the float \code{0.1} that results from conversion is therefore not \code{"1/10"} but        
\begin{Schunk}
\begin{Soutput}
[1] "3602879701896397/36028797018963968"
\end{Soutput}
\end{Schunk}
When the above is converted back into floating-point numbers it is \code{0.1}, so the inaccuracy is hidden. Conversely, if we supply the number as \code{"1/10"} from the beginning and then turn that into a floating-point number, we also get \code{0.1} but this time from the rational number \code{"1/10"} without inaccuracy. For that reason it would be best to have the data to be checked available in this character representation already but we are aware that this is most likely not going to happen. Going forward, whenever numbers are represented as characters, we utilize rational arithmetic; ``normal'' numbers refer to floating point.


\subsection{Checking for separation: A general linear program for separation} 
\label{sec:genlin}
Extending the idea of \cite{polytomous:Albert+Anderson:1984,konis2007} and similar to \cite{Fienberg+Rinaldo:2012}, \cite{Sablicaetal} suggest a general, versatile check for separation that can be used for all categorical response model and data situations. The versatility stems from it being based solely on the separation theory for structure vectors.

The following linear program can be used to check separation for all data cases:
\begin{equation}
 \begin{aligned}
  \text{min:}\ y  \\
\textup{such that}\quad S^{\top}u   &= \mathbf{0}_{p}\\
                    u + y \mathbf{1}_{\nS} &\geq \mathbf{1}_{\nS},\\
                     y & \geq 0
\end{aligned}
\label{lp1}
\end{equation}
with $\mathbf{1}_{\nS}$ and $\mathbf{0}_{p}$ the $\nS$ or $p$-dimensional vector of $1$s or $0$s respectively. If the solution is exactly equal to $0$ then the dual is feasible and we have overlap. If it is not equal to $0$ the dual has no feasible solution and we have separation. This program  can be solved globally with polynomial complexity. The distinguishing feature of the linear program in Equation~\ref{lp1} is that it can be used for any latent polyhedral categorical response model, as long as $S$ is setup correctly and that it works in every case (including rank-deficiency) due to this program being the Farkas/dual-style formulation of overlap versus separation. It therefore gives back the canonical certificate. See the description in \cite{Sablicaetal}. 
 
The functions in \pkg{divoRce} that make use of linear program in Equation~\ref{lp1} are the generic function \fct{check\_separation} with argument \code{quick = FALSE} and \fct{check\_overlap} with argument \code{quick = FALSE}.

\fct{check\_overlap} also allows to check for overlap sequentially (\code{sequential = TRUE}) or in parallel (utilizing multiple cores, \code{parallel = TRUE}) on subsets of the data. The number of subsets can be specified with \code{nss} and the number of cores for parallel execution with \code{nc}. This is expected to be quicker if overlap exists, because as long as the subset spans the full $\R^p$ any overlap in a subset means overlap for the full data. Hence if there is no reason to suspect separation, one can quickly confirm overlap with this function. Note that if some subsets do not span the full $\R^p$, we need to run the full linear program at the end to determine whether overlap exists in the full data, so in that case sequential or parallel might not be more efficient. We include an argument \code{shuffle} (defaults to \code{TRUE}) that shuffles the data before subsets are built to make it less likely that the former will happen.  

All of these functions return either \code{TRUE} or \code{FALSE}.

To sum up, we have these options:
\begin{description}
\item[\code{check\_separation(..., quick = FALSE)}:] Check function for separation via linear program Equation~\ref{lp1}.
\item[\code{check\_overlap(..., quick = FALSE)}:] Check function for overlap  via linear program Equation~\ref{lp1}.
\item[\code{check\_overlap(..., parallel = TRUE, nc, nss, shuffle)}:] Check function for overlap via linear program in Equation~\ref{lp1} by checking \code{nss} subsets of the data in parallel on \code{nc} cores.
\item[\code{check\_overlap(..., sequential = TRUE, nss, shuffle)}:] Check function for overlap via linear program in Equation~\ref{lp1} by sequentially checking \code{nss} subsets of the data.
\end{description}  
As default solver we use \fct{rcdd::lpcdd} but other solvers can be chosen as well if floating-point arithmetic suffices.

\subsubsection{Solvers}
\label{sec:solvers}
To solve the linear programs of Equation~\ref{lp1} and Equation~\ref{lp2}, users can specify different backends and solvers to be used by the functions \fct{check\_separation}, \fct{check\_overlap}, \fct{diagnose\_separation} and \fct{separation\_columns}. This is done via the \code{backend} and \code{solver} arguments.

We support two backends: \code{"rcdd"} uses \pkg{rcdd}'s \fct{lpcdd} which has implemented two solvers, \code{"DualSimplex"} (default) and \code{"CrissCross"}, and \code{"ROI"} which provides access to many solvers for linear programs via plugins to \pkg{ROI} \citep{ROIpap}, including \pkg{lpsolve} \citep{lpsolve, lpsolveROI}, \pkg{glpk} \citep{glpk, glpkROI}, \pkg{hiGHs} \citep{highs, highsROI} and others. If arbitrary precision is sought, then \pkg{rcdd} has to be used as backend and one can choose one of the two solvers. If floating-point arithmetic is sufficient, one can use both \pkg{rcdd} and \pkg{ROI}. If no backend is specified the functions default to using \pkg{rcdd} as the backend with \code{"DualSimplex"} as the solver. 

Different solvers and backends are specified for the mentioned functions in the following way (illustrated for \pkg{ROI} as backend and \pkg{hiGHs} as solver)
\begin{Schunk}
\begin{Sinput}
> check_separation(y, X, backend = "ROI", solver = "highs")
\end{Sinput}
\end{Schunk}

Naturally the backend and solver used have a bearing on the performance. We default to using \pkg{rcdd} as backend because we use a number of functions from it and not just the linear program solvers. However, the \pkg{rcdd} implementation for solving a linear program \fct{lpcdd} can be slow for larger problems (e.g., large number of rows in $S$). For medium sized to larger problems we recommend to use \pkg{ROI} as backend with an optimized solver, for example \code{solver = "highs"}. If rational arithmetic is desired, the only option is using \pkg{rcdd} as backend.

\subsection{Linearities: Distinguishing types of separation}
\label{sec:linearities}
A special feature of our diagnostic suite is to not only check whether there is separation but to reliably detect which of the three cases (overlap, complete or quasi-complete separation) exist in the structure vectors. Numerically, this is not easy; distinguishing quasi-complete separation from the other cases means checking equality in a linear program and that can be difficult in floating-point arithmetic \citep{konis2007}. To do this properly one requires arbitrary precision as described in Section~\ref{sec:rational}.

For diagnosing cases of separation we are able to work with the fact that in case of any separation, the structural vectors generate a recession cone. \cite{geyer2009likelihood} uses this to calculate generic directions of recession for the MLE in the binary case and we extend this idea to nominal and ordinal responses by checking for linearities in the finitely generated polyhedral cone of the structure vectors \citep[Lemmas~1, 4 in][]{Sablicaetal}.

In \pkg{divoRce} we provide a mid-level (non-generic) function \fct{linearities} that allows to check for the existence of linearities by calculating all the nonlinearity generators in the (negative) structure vector matrix that can be made linear without changing the cone. It takes as arguments either path (a), the response vector \code{y}, a design matrix \code{X}, and the link construction via \code{model}, or via path (b) the structure vector matrix \code{S}, and then calculates which negative structure vectors are linearities. Note we do not have a generic for \fct{linearities} as it is mostly a helper function used in \fct{diagnose\_separation}.

The function \fct{linearities} employs computational geometry based on \fct{linearity} in \pkg{rcdd} and operates on the vertex representation (V-rep) of the polyhedral cone, which means the convex hull of points (vertices) and the generating vectors (rays). In V-rep, a linearity is the largest linear subspace contained in the cone. In terms of the model, this would be the smallest, non-redundant linear model (affine set) spanned by the structure vectors. Therefore a structure vector being a linearity means that it is in the ``well-behaved'' (overlap) set. 

It holds that if the set of linearities is empty (so the number of linearities is $0$), we have complete separation. If there is overlap, the number of linearities is equal to the number of structure vectors, so every row in $S$ is a linearity and we have $\nS$ linearities. If neither happens, we have quasi-complete separation \citep[the number of linearities is between $0$ and $\nS$, Lemma~4;][]{Sablicaetal}. 

The following example illustrates the idea of using linearities for distinguishing cases of separation. We use binary overlap data
\begin{Schunk}
\begin{Sinput}
> data("ovldat1")
\end{Sinput}
\end{Schunk}
to which we fit an additive logistic regression model
\begin{Schunk}
\begin{Sinput}
> m1 <- glm(y ~ x1 + x2, family = "binomial", data = ovldat1)
\end{Sinput}
\end{Schunk}
Extracting the responses and the design matrix from the model object directly and supplying them via path (a), we can find the linearities in the cone due to $S$ via, 
\begin{Schunk}
\begin{Sinput}
> louto <- linearities(m1$y, model.matrix(m1), model = "b")
> louto$index
\end{Sinput}
\begin{Soutput}
 [1]  1  2  3  4  5  6  7  8  9 10 11 12 13 14 15 16 17 18 19 20 21 22 23 24 25
\end{Soutput}
\end{Schunk}
In this example the rows $1$ through $25$ of $S$ are all linearities. Since this is also the number of rows in $S$, we have overlap. 

Here is how this would look like for complete separation:
\begin{Schunk}
\begin{Sinput}
> data("csepdat1")
> m2 <- glm(y ~ x1 + x2, family = "binomial", data = csepdat1)
\end{Sinput}
\begin{Soutput}
Warning: glm.fit: algorithm did not converge
\end{Soutput}
\begin{Soutput}
Warning: glm.fit: fitted probabilities numerically 0 or 1 occurred
\end{Soutput}
\begin{Sinput}
> loutcs <- linearities(m2$y, model.matrix(m2), model = "b")
> loutcs$index
\end{Sinput}
\begin{Soutput}
integer(0)
\end{Soutput}
\end{Schunk}
Now \fct{linearities} returns the empty set which is indicative of complete separation.  

Finally here is how this would look like for quasi-complete separation:  
\begin{Schunk}
\begin{Sinput}
> data("csepdat1")
> m3 <- glm(y ~ x1 + x2, family = "binomial", data = qcsepdat1)
\end{Sinput}
\begin{Soutput}
Warning: glm.fit: algorithm did not converge
\end{Soutput}
\begin{Soutput}
Warning: glm.fit: fitted probabilities numerically 0 or 1 occurred
\end{Soutput}
\begin{Sinput}
> loutcs <- linearities(m3$y, model.matrix(m3), model = "b")
> loutcs$index
\end{Sinput}
\begin{Soutput}
[1] 3 4 5
\end{Soutput}
\end{Schunk}
Here our linearity set comprises three structure vectors, the rows $3, 4, 5$.  Since this is not an empty set but of cardinality less than $\nS=25$, we have quasi-complete separation.    

For nominal and ordinal models we can analogously check by making use of \code{y}, \code{X} or \code{S} matrices as defined in Section~\ref{sec:genlin} and conduct the above steps.

\subsubsection{Identification of separation rows}
We can use linearities in the structure vectors as described above for identification of which rows in the design matrix or structure vector matrix determine the separation phenomenon, if any. For this we can find the set of rows in the design matrix/structure vector matrix who have at least one structure vector that is not a linearity; any row that is in that set is ``offending'' i.e., contributes to separation. This is again an application of Lemma~4 in \citet{Sablicaetal}. For this we have the generic \fct{separation\_rows}. In case of a supplied $y$ and $X$ this returns the submatrix of the $(X,y)$-tuple that features all the rows in $X$ that have at least one associated structure vector in $S$ that is not a linearity (\code{$offrows}, there can be more than one such structure vector for a row in $X$), as well as the index of these rows in $X$ (\code{$index}). The former is meant for inspection and the latter for subsetting/adressing the rows in the data themselves. For a supplied structure vector matrix $S$ it simply returns all the structure vectors that are not linearities as well as their row indices in $S$.

Here is an example:
\begin{Schunk}
\begin{Sinput}
> seprows1 <- separation_rows(m3$y, model.matrix(m3), model = "b")
> seprows1
\end{Sinput}
\begin{Soutput}
$offrows
   (Intercept)        x1          x2 y
1            1  1.000000  3.00000000 0
2            1  2.000000  0.00000000 0
6            1  4.000000  0.00000000 1
7            1  5.000000  2.00000000 1
8            1  6.000000  7.00000000 1
9            1 10.000000  3.00000000 1
10           1 11.000000  4.00000000 1
11           1  7.923643 -0.71695576 1
12           1  5.841141  5.55543886 1
13           1  6.606504  0.80400287 1
14           1  9.176108  4.41846461 1
15           1  6.650658  5.92632068 1
16           1 10.345943 -0.60978838 1
17           1  9.516725  0.07964299 1
18           1 11.581141  4.66743639 1
19           1 10.809490 -0.24339052 1
20           1  8.031372  1.19162292 1
21           1  1.923957  9.10082960 0
22           1  2.943028 10.74461711 0
23           1  1.455745  7.23534985 0
24           1  2.385796  8.15634364 0
25           1  2.450655  2.69138426 0

$index
 [1]  1  2  6  7  8  9 10 11 12 13 14 15 16 17 18 19 20 21 22 23 24 25
\end{Soutput}
\end{Schunk}
Since our linearity set comprised the structure vectors for observations $3, 4, 5$, we now have all other rows in $S$ associated with separation. This approach also works with the respective function for the nominal and ordinal data/model combinations.

\subsection{Recession cone and its dimension}
\label{sec:rccone}
With the geometry of separation that we use to diagnose the cases, we can also provide a feature to calculate the recession cone as well as its dimension. Relating the separation cases to the dimension of the recession cone goes back to \cite{polytomous:Kaufmann:1988}. For full-rank $S$, complete separation implies a dimension of $p$ for the recession cone and overlap implies a dimensionality of $0$. Quasi-complete separation is the most interesting one and implies a dimension between $0$ and $p$.

The mathematical basis or finding the dimension of the recession cone is Theorem~7 of \cite{geyer2009likelihood} and is explicated for our case as Theorem~1 and Lemma~4 in \cite{Sablicaetal} or Theorem~4 of \cite{geyer2009likelihood}. They all imply that separation only exists if the linearities do not span the whole space (so, if there is no overlap). The cone generated by the structure vectors, $K$, can be decomposed into lineality space $L$ containing the linearities and a pointed cone $P$. The space spanned by the linearities in $L$ is orthogonal to the directions of recession $R$ and their dimensions sum up to the dimension of the whole cone $K$. We use this to get the dimensionality of the recession cone as $\text{dim}(R)=\text{dim}(P)=\text{dim}(K)-\text{dim}(L)=\text{rank}(S)-\text{dim}(L)$. For this we calculate the column rank for the submatrix of $S$ that comprises the structure vectors that are linearities (our $\text{dim}(L)$) and subtract it from the rank of $S$. We also return the structure vectors that are not in $L$, i.e., vectors that are non-lineality generators of $K$. 

To calculate both the cone and its dimension \pkg{divoRce} contains the generic \fct{recession\_cone} function. To illustrate for the binary case once again, we use the quasi-complete data example from the previous section where $p$ is $3$ (the variables \code{x1} and \code{x2} and the intercept). 
\begin{Schunk}
\begin{Sinput}
> recc1 <- recession_cone(m3$y, model.matrix(m3), model = "b")
> recc1$reccdim
\end{Sinput}
\begin{Soutput}
[1] 1
\end{Soutput}
\end{Schunk}
We see that the dimension of the recession cone is $\text{dim}(R)=1$. Hence the recession direction is a one-dimensional ray starting at the origin. This is quasi-complete separation as $0 < \text{dim}(R)=1 < p = 3$.

The structure vectors in the recession cone itself are given by
\begin{Schunk}
\begin{Sinput}
> recc1$cone
\end{Sinput}
\begin{Soutput}
     1::(Intercept)     1::x1        1::x2
1.1              -1 -1.000000  -3.00000000
2.1              -1 -2.000000   0.00000000
6.1               1  4.000000   0.00000000
7.1               1  5.000000   2.00000000
8.1               1  6.000000   7.00000000
9.1               1 10.000000   3.00000000
10.1              1 11.000000   4.00000000
11.1              1  7.923643  -0.71695576
12.1              1  5.841141   5.55543886
13.1              1  6.606504   0.80400287
14.1              1  9.176108   4.41846461
15.1              1  6.650658   5.92632068
16.1              1 10.345943  -0.60978838
17.1              1  9.516725   0.07964299
18.1              1 11.581141   4.66743639
19.1              1 10.809490  -0.24339052
20.1              1  8.031372   1.19162292
21.1             -1 -1.923957  -9.10082960
22.1             -1 -2.943028 -10.74461711
23.1             -1 -1.455745  -7.23534985
24.1             -1 -2.385796  -8.15634364
25.1             -1 -2.450655  -2.69138426
\end{Soutput}
\end{Schunk}
consisting of all observations but number $3, 4, 5$, the linearities identified earlier. For all other latent polyhedral CRM, it works the same way.

\subsection{Detection of  separation columns: Another linear program for separation}
\label{sec:genlin2}
A further feature of our diagnostic suite is to detect the columns of the design matrix/structure matrix that are the culprits behind separation if it exists. For a structure vector matrix $S$, we solve the following linear program with a similar rationale as Equation~\ref{lp1}:
\begin{equation}
\begin{aligned}
\max_\beta:\quad \mathbf{1}_{\nS}^{\top}S\beta\\
\text{such that}\quad -S\beta&\leq \mathbf{0}_{\nS},\\
                       -\beta&\leq\mathbf{1}_{p},\\
                        \beta&\leq\mathbf{1}_{p}.
    \end{aligned}    
\label{lp2}    
\end{equation}
with $\bm{1}_p$ and $\bm{1}_{\nS}$  denoting the $p$ and $\nS$-dimensional vector of ones, respectively. The idea behind this linear program is this: Say we have separation for a column of structure vector matrix $S$. Then maximizing the sum of the product of each value in that column of $S$ with the corresponding $\beta_b$ over $\beta$ would send $\beta_b$ towards $\pm \infty$ as all values in that column point towards the same direction for the same category; the $\beta_b$ can be made infinitely large. Conversely, if there is overlap then at least one entry in the column of $S$ has the ``wrong'' sign in the sum of the product and we add at least one $\infty$ with the opposite sign to the sum. Thus in case of overlap the solution for these $\beta_b$ must be $0$. We further set the boundaries of $\beta_b$ to $-1$ and $1$, so $\infty$ would correspond to the lower or upper bound of the constraint. Hence if the value returned is $\beta_b\neq 0$ we have separation for column $b$ and if not ($\beta_b=0$) we have overlap. 

We have implemented this in the generic \fct{separation\_columns}. This functions supports choosing backend and solver with the arguments \code{backend} and \code{solver} as described in Section~\ref{sec:solvers}.

Let us illustrate by returning to the example from the introduction, the trips to Lake Somerville. We saw the coefficient values for ``userfeeyes'' to be quite large, which points towards this being responsible fro separation. We now run a proper check with the low-level function and use \pkg{ROI} as backend with the ``lpsolve''-solver. 

\begin{Schunk}
\begin{Sinput}
> ds1 <- separation_columns(m.recdem$y, model.matrix(m.recdem),
+     model = "b", backend = "ROI", solver = "lpsolve")
> ds1
\end{Sinput}
\begin{Soutput}
$ls
[1]  0.000000e+00  3.125098e-16  1.000000e+00 -4.535668e-16

$offcols
[1] "userfeeyes"

$colnrs
[1] 3

$separated
[1] FALSE FALSE  TRUE FALSE
\end{Soutput}
\end{Schunk}
The function returns the solution of the primal linear program in \code{$ls}, a boolean of whether each column shows separation (\code{$separated}) and the index of the columns associated with separation (\code{$colnrs}). The column name of the columns with \code{separated} equal to \code{TRUE} are given as \code{$offcols}. Here it is the third column of the design matrix, where $\beta_3=-1$ and all others are essentially zero (in floating-point precision). Thus the third column in the design matrix exhibits separation, but the others do not. 

The linear program in Equation~\ref{lp2} can typically be solved faster than the one in Equation~\ref{lp1} as it checks in coefficient space if the primal problem is bounded for any column, so only needs to check $p$ variables instead of $\nS$, which is usually much less. We can use this as a basis for a quick check for separation/overlap that simply checks whether any $\beta_b \neq 0$ and, if so, concludes separation. While quicker, it does not provide the same reliability as the full linear program based on the dual in Equation~\ref{lp1}, for example in case of rank deficiency. In the generics \fct{check\_separation} and \fct{check\_overlap} this can be chosen with the argument \code{quick = TRUE}. This also works for \code{parallel = TRUE} and \code{sequential = TRUE} in \fct{check\_overlap} but is unreliable in case of non-full rank.

This function family can also be used to showcase why using rational arithmetic can be important. Strictly speaking the values we output above in \code{$ls} are all not exactly equal to $0$ which is due to floating-point arithmetic. So when utilizing floating-point, one needs to take a tolerance into account to get exact zeros/ones, e.g., by doing something like \code{abs(ds1$ls) < tolerance} to define an exact $0$.  It is therefore possible that a linear program will output a value which might be wrongly interpreted as indicative of separation or not if the tolerance is not exceeded. For example, the check underlying \fct{detect\_infinite\_estimates} in \pkg{detectseparation} uses a tolerance of $1 \times 10^{-4}$ below which a value is interpreted as $0$ and finiteness of a coefficient is concluded. If we set the tolerance to \code{1e-16} we would conclude all coefficients but the one for the intercept are infinite. 
\begin{Schunk}
\begin{Sinput}
> glm(TripYes ~ costS + userfee + costC, family = binomial, data = recdem,
+     method = detectseparation::detect_separation, tolerance = 1e-16)
\end{Sinput}
\begin{Soutput}
Implementation: ROI | Solver: lpsolve 
Separation: TRUE 
Existence of maximum likelihood estimates
(Intercept)       costS  userfeeyes       costC 
          0         Inf         Inf        -Inf 
0: finite value, Inf: infinity, -Inf: -infinity
\end{Soutput}
\end{Schunk}
In contrast, if we use \code{divoRce::detect\_sepcols} and set \code{rational = TRUE}, we get 
\begin{Schunk}
\begin{Sinput}
> ds2 <- separation_columns(m.recdem$y, model.matrix(m.recdem),
+     model = "b", rational = TRUE)
> ds2$ls
\end{Sinput}
\begin{Soutput}
[1] 0 0 1 0
\end{Soutput}
\end{Schunk}
so the values are exact and there is no possible error. With the $S$ from Section~\ref{sec:genlin} this also works for nominal and ordinal responses.

\subsection{Full diagnostic convenience function}
\label{sec:fulldiag}
All the diagnostic pieces illustrated above can be addressed directly with the respective specialized generic and that is the most efficient way to get individual information such as whether separation exists or not. We do believe that many users will also want an overview of all of the main aspects at the same time as a one-stop solution for separation diagnostics. For this we implemented the \code{diagnose\_separation} generic and its methods for pre- and post-fit as described in Section~\ref{sec:design}. It utilizes the UI described earlier. This function calls many of the functions we described previously and combines their outputs to single object of class \class{sepmod}. We have implemented a S3 \code{print} method for them. This \code{print.sepmod} has as an additional argument \code{info} that prints either a minimalist summary of the separation situation (\code{info = "minimal"}, also the default) and \code{info = "full"} that gives all available information.

To illustrate let us use the data set of \cite{polytomous:Silvapulle:1981}.

\begin{Schunk}
\begin{Sinput}
> data("Silvapulle")
\end{Sinput}
\end{Schunk}
We first fit an additive model binary regression mode with a cloglog link and note we get the \proglang{R} warning of fitted probabilities of 0 or 1 occurred.  
\begin{Schunk}
\begin{Sinput}
> mSilvA <- glm(case ~ sex + ghq, data = Silvapulle, family = binomial(link = "cloglog"))
\end{Sinput}
\begin{Soutput}
Warning: glm.fit: fitted probabilities numerically 0 or 1 occurred
\end{Soutput}
\end{Schunk}
We might be tempted to think this is separated, but we actually have overlap here:
\begin{Schunk}
\begin{Sinput}
> dsA <- diagnose_separation(mSilvA, rational = TRUE, backend = "rcdd",
+     solver = "CrissCross")
> print(dsA, info = "minimal")
\end{Sinput}
\begin{Soutput}
Separation Check for Model: 
glm(formula = case ~ sex + ghq, family = binomial(link = "cloglog"), 
    data = Silvapulle)
Check Family: Binary Model 

Separation: FALSE 
Separation Type: Overlap 
\end{Soutput}
\end{Schunk}

When fitting an interaction model we get the \proglang{R} warning of fitted probabilities of 0 or 1 occurred again.  
\begin{Schunk}
\begin{Sinput}
> mSilvM <- update(mSilvA, case ~ sex * ghq)
\end{Sinput}
\begin{Soutput}
Warning: glm.fit: fitted probabilities numerically 0 or 1 occurred
\end{Soutput}
\end{Schunk}
Running the full diagnostic suite
\begin{Schunk}
\begin{Sinput}
> dsM <- diagnose_separation(mSilvM, rational = TRUE, model = "b",
+     backend = "rcdd", solver = "CrissCross")
> print(dsM, info = "full")
\end{Sinput}
\begin{Soutput}
Separation Check for Model: 
glm(formula = case ~ sex + ghq + sex:ghq, family = binomial(link = "cloglog"), 
    data = Silvapulle)
Check Family: Binary Model 

Separation: TRUE 
Separation Type: Quasi-Complete Separation 

Dimension of Recession Cone: 1 
Number of Offending Columns: 4 
Number of Offending Rows: 33 
\end{Soutput}
\end{Schunk}
we see that we do have separation, specifically quasi-complete separation. It effects all four coefficients due to $33$ of the $120$ observations. The slots \code{dsM$offrows} and \code{dsM$offcols} would return the rows affected by separation and the separation columns respectively (as also returned by \fct{separation\_rows} and \fct{separation\_columns}). 

The drawback of using this function is that this can take a long while due to solving linear programs as well as calculating linearities, separation rows, separation columns and recession cone in one go, so if execution speed is a factor, it is better to rely on the specialized functions. 

\section{Further illustrations}
In this section we illustrate the usage of \pkg{divoRce} functionality with three situations for which one can currently use our software: binary responses, unordered categorical responses and a baseline-category link model and ordinal responses and cumulative link models. 
 For these examples, we will also give timings of how long it takes to do the calculations in different setups with \proglang{R} 4.6.0 on a Lenovo X13 Yoga Notebook with 11th Gen Intel Core i5-1135G7 @ 2.40GHz processor and Linux Mint 22.3 for \pkg{divoRce} 0.13-0.

\subsection{Binary responses}
The data set used by \cite{Ruschetal:2013} features a target variable that is binary encoding whether a person took part in the US general election in 2004 (``gen04''). It consists of $n=6544$ eligible voters from Ohio and contains $83$ variables that \cite{Ruschetal:2013} used for predicting individual turnout probability with the goal of using it for voter targeting. In the section where they developed voter profiles for targeting, they encounter a separation phenomenon.

The voter profiles employ a number of predictors that we reuse here, namely whether the individual voted in the general elections from 2000 to 2003 (``genXX'') and the presidential primary election in 2004 (``ppp04''), the percentage of elections attended out of all that they were eligible to (``percentAttended''), their age in days (as a cubic spine), the individuals rank in the household (with rank of 1 meaning ``head of the household'') and the variable ``partyMix'' specifying how the household is made up of in terms of party affiliation (for example, ``allD'' or ``noneR'' would mean all members of the household are registered Democrats or none of the members of the household are registered Republicans respectively; the two are not the same). We opt for fitting a probit model for the target variable with the predictors in an additive specification. 
\begin{Schunk}
\begin{Sinput}
> library("splines")
> data("ohiovoters")
> m1 <- stats::glm(gen04 ~ gen00 + gen01 + gen02 + gen03 + ppp04 +
+     bs(ageD) + householdRank + percentAttended + partyMix, data = ohiovoters,
+     family = binomial(link = "probit"))
> summary(m1)
\end{Sinput}
\begin{Soutput}

Call:
stats::glm(formula = gen04 ~ gen00 + gen01 + gen02 + gen03 + 
    ppp04 + bs(ageD) + householdRank + percentAttended + partyMix, 
    family = binomial(link = "probit"), data = ohiovoters)

Coefficients:
                       Estimate Std. Error z value Pr(>|z|)    
(Intercept)            -0.99247    0.13380  -7.417 1.19e-13 ***
gen00                   0.79798    0.05417  14.732  < 2e-16 ***
gen01                  -0.24871    0.05792  -4.294 1.76e-05 ***
gen02                   0.64226    0.05605  11.460  < 2e-16 ***
gen03                   0.53664    0.05934   9.044  < 2e-16 ***
ppp04                   0.38361    0.07097   5.405 6.48e-08 ***
bs(ageD)1               0.18456    0.30062   0.614 0.539254    
bs(ageD)2               0.71454    0.28964   2.467 0.013626 *  
bs(ageD)3              -0.71054    0.37949  -1.872 0.061154 .  
householdRank2          0.25260    0.04854   5.204 1.95e-07 ***
householdRank3+        -0.10211    0.09247  -1.104 0.269497    
percentAttended        -0.72054    0.21241  -3.392 0.000693 ***
partyMixallR            0.34819    0.08224   4.234 2.30e-05 ***
partyMixallRorDorLegal  0.91925    0.16854   5.454 4.92e-08 ***
partyMixnoneD           1.05434    0.09893  10.657  < 2e-16 ***
partyMixnoneR           0.99887    0.09726  10.270  < 2e-16 ***
partyMixnoneRorD        0.63706    0.08783   7.253 4.07e-13 ***
partyMixonlyRorD        0.24065    0.12285   1.959 0.050130 .  
partyMixunknown        -6.42817   31.48922  -0.204 0.838245    
---
Signif. codes:  0 '***' 0.001 '**' 0.01 '*' 0.05 '.' 0.1 ' ' 1

(Dispersion parameter for binomial family taken to be 1)

    Null deviance: 8036.6  on 6543  degrees of freedom
Residual deviance: 4704.8  on 6525  degrees of freedom
AIC: 4742.8

Number of Fisher Scoring iterations: 16
\end{Soutput}
\end{Schunk}
The \fct{glm} function does not give a warning and the output does not appear too alarming either at first glance. Upon closer inspection, the coefficient for \code{partyMixunknown} of -6.43 might be considered rather low, but the $p-$~value associated with the standard Wald test for the coefficient is around $0.84$, so likely to be interpreted as ``not significant'' and thus one might conclude that this variable is not relevant for predicting the target. We can use \pkg{divoRce} diagnostics to see that this would be a mistake (note we are in a post-fit situation).

We run our separation diagnostics convenience function. 
\begin{Schunk}
\begin{Sinput}
> d1 <- diagnose_separation(m1)
> print(d1, "full")
\end{Sinput}
\begin{Soutput}
Separation Check for Model: 
stats::glm(formula = gen04 ~ gen00 + gen01 + gen02 + gen03 + 
    ppp04 + bs(ageD) + householdRank + percentAttended + partyMix, 
    family = binomial(link = "probit"), data = ohiovoters)
Check Family: Binary Model 

Separation: TRUE 
Separation Type: Quasi-Complete Separation 

Dimension of Recession Cone: 1 
Number of Offending Columns: 1 
Number of Offending Rows: 757 
\end{Soutput}
\end{Schunk}
We have quasi-complete separation and $757$ rows of $S$ (here corresponding to observations) are affected by this. Rows affected can be be extracted via \code{d1$offrows} and the affected design matrix column with \code{d1$offcols}.
\begin{Schunk}
\begin{Sinput}
> d1$offcols
\end{Sinput}
\begin{Soutput}
[1] "partyMixunknown"
\end{Soutput}
\end{Schunk}
The responsible columns in the design matrix is therefore \code{partyMixunknown}: Everyone that has ``partyMix'' equal to ``unknown'' has not turned out to vote in the general election of 2004, which can also be seen in the subsequent contingency table:
\begin{Schunk}
\begin{Sinput}
> with(ohiovoters, table(gen04, partyMix))
\end{Sinput}
\begin{Soutput}
     partyMix
gen04 allD allR allRorDorLegal noneD noneR noneRorD onlyRorD unknown
    0  157   92             24    93    96      736       33     757
    1  634  646            109   721   660     1600      186       0
\end{Soutput}
\end{Schunk}
This means the ``partyMix'' variable is in fact highly important for prediction and with it we can increase predictive accuracy by almost $12\%$ (the fraction of people with unknown party mix in the data which we would be able to predict perfectly). Indeed using \citet{kosmidis2021} bias-reduced estimation procedure from \pkg{brglm2} \citep{brglm2} 
\begin{Schunk}
\begin{Sinput}
> m1a <- update(m1, method = brglm2::brglmFit)
> summary(m1a)
\end{Sinput}
\begin{Soutput}

Call:
stats::glm(formula = gen04 ~ gen00 + gen01 + gen02 + gen03 + 
    ppp04 + bs(ageD) + householdRank + percentAttended + partyMix, 
    family = binomial(link = "probit"), data = ohiovoters, method = brglm2::brglmFit)

Deviance Residuals: 
    Min       1Q   Median       3Q      Max  
-3.1915  -0.0230   0.3384   0.5947   1.9263  

Coefficients:
                       Estimate Std. Error z value Pr(>|z|)    
(Intercept)            -0.98589    0.13348  -7.386 1.51e-13 ***
gen00                   0.79522    0.05411  14.697  < 2e-16 ***
gen01                  -0.24728    0.05777  -4.280 1.87e-05 ***
gen02                   0.63935    0.05592  11.432  < 2e-16 ***
gen03                   0.53409    0.05918   9.025  < 2e-16 ***
ppp04                   0.38064    0.07058   5.393 6.92e-08 ***
bs(ageD)1               0.18109    0.30017   0.603 0.546310    
bs(ageD)2               0.71719    0.28894   2.482 0.013059 *  
bs(ageD)3              -0.72005    0.37883  -1.901 0.057337 .  
householdRank2          0.25129    0.04838   5.194 2.05e-07 ***
householdRank3+        -0.10138    0.09233  -1.098 0.272232    
percentAttended        -0.71712    0.21167  -3.388 0.000704 ***
partyMixallR            0.34609    0.08209   4.216 2.49e-05 ***
partyMixallRorDorLegal  0.90546    0.16776   5.397 6.76e-08 ***
partyMixnoneD           1.04680    0.09855  10.622  < 2e-16 ***
partyMixnoneR           0.99157    0.09691  10.232  < 2e-16 ***
partyMixnoneRorD        0.63277    0.08753   7.229 4.86e-13 ***
partyMixonlyRorD        0.23551    0.12251   1.922 0.054560 .  
partyMixunknown        -4.01848    0.44733  -8.983  < 2e-16 ***
---
Signif. codes:  0 '***' 0.001 '**' 0.01 '*' 0.05 '.' 0.1 ' ' 1

(Dispersion parameter for binomial family taken to be 1)

    Null deviance: 8036.6  on 6543  degrees of freedom
Residual deviance: 4705.8  on 6525  degrees of freedom
AIC:  4743.8

Type of estimator: AS_mixed (mixed bias-reducing adjusted score equations)
Number of Fisher Scoring iterations: 5
\end{Soutput}
\end{Schunk}
we find that the \code{partyMixunknown} is a strong predictor in the model. 

That correct diagnosis is important can be seen by other model specifications that give results that might look like separation but are not. To illustrate, we run the following model which amongst other effects specifies a three-way interaction between \code{dontPhone}, \code{percentAttended} and \code{householdRank}.   
\begin{Schunk}
\begin{Sinput}
> m2 <- glm(ppp04 ~ gen00 + dontPhone * householdRank * percentAttended +
+     ageD, data = ohiovoters, family = binomial("logit"))
> summary(m2)
\end{Sinput}
\begin{Soutput}

Call:
glm(formula = ppp04 ~ gen00 + dontPhone * householdRank * percentAttended + 
    ageD, family = binomial("logit"), data = ohiovoters)

Coefficients:
                                             Estimate Std. Error z value
(Intercept)                                -2.971e+00  1.827e-01 -16.264
gen00                                       4.915e-01  9.932e-02   4.949
dontPhoneU                                 -1.826e-01  6.225e-01  -0.293
dontPhoneY                                 -4.013e-02  1.976e-01  -0.203
householdRank2                              1.232e-01  2.258e-01   0.545
householdRank3+                            -1.608e+00  5.206e-01  -3.090
percentAttended                             8.245e+00  4.604e-01  17.910
ageD                                       -1.302e-06  7.093e-06  -0.184
dontPhoneU:householdRank2                  -9.105e-01  1.203e+00  -0.757
dontPhoneY:householdRank2                   7.062e-02  3.166e-01   0.223
dontPhoneU:householdRank3+                  1.741e+00  1.326e+00   1.313
dontPhoneY:householdRank3+                  3.993e-01  6.598e-01   0.605
dontPhoneU:percentAttended                 -7.365e-01  1.729e+00  -0.426
dontPhoneY:percentAttended                  4.740e-02  6.193e-01   0.077
householdRank2:percentAttended             -2.214e-01  7.380e-01  -0.300
householdRank3+:percentAttended             5.113e+00  1.840e+00   2.779
dontPhoneU:householdRank2:percentAttended   2.073e+00  3.093e+00   0.670
dontPhoneY:householdRank2:percentAttended  -1.872e-01  9.906e-01  -0.189
dontPhoneU:householdRank3+:percentAttended -3.831e+00  4.057e+00  -0.944
dontPhoneY:householdRank3+:percentAttended -1.989e+00  2.311e+00  -0.860
                                           Pr(>|z|)    
(Intercept)                                 < 2e-16 ***
gen00                                      7.47e-07 ***
dontPhoneU                                  0.76928    
dontPhoneY                                  0.83904    
householdRank2                              0.58549    
householdRank3+                             0.00200 ** 
percentAttended                             < 2e-16 ***
ageD                                        0.85439    
dontPhoneU:householdRank2                   0.44897    
dontPhoneY:householdRank2                   0.82349    
dontPhoneU:householdRank3+                  0.18917    
dontPhoneY:householdRank3+                  0.54503    
dontPhoneU:percentAttended                  0.67023    
dontPhoneY:percentAttended                  0.93899    
householdRank2:percentAttended              0.76416    
householdRank3+:percentAttended             0.00546 ** 
dontPhoneU:householdRank2:percentAttended   0.50276    
dontPhoneY:householdRank2:percentAttended   0.85010    
dontPhoneU:householdRank3+:percentAttended  0.34500    
dontPhoneY:householdRank3+:percentAttended  0.38957    
---
Signif. codes:  0 '***' 0.001 '**' 0.01 '*' 0.05 '.' 0.1 ' ' 1

(Dispersion parameter for binomial family taken to be 1)

    Null deviance: 8907.2  on 6543  degrees of freedom
Residual deviance: 6423.1  on 6524  degrees of freedom
AIC: 6463.1

Number of Fisher Scoring iterations: 5
\end{Soutput}
\end{Schunk}
Here the coefficients for the partial effects that we obtain for \code{percentAttended} and the interaction term \code{householdRank3+:percentAttended} seem high and one might be tempted to take the effect size at face value to be of similar magnitude as for the separation variable in the prior model (which is wrong since the link family distribution differs; in general a logit coefficient is roughly $1.6$ times larger than the corresponding probit coefficient). Let us do a sequential quick check for overlap to see what is going on:
\begin{Schunk}
\begin{Sinput}
> check_overlap(m2$y, model.matrix(m2), model = "bcl", quick = TRUE,
+     sequential = TRUE, backend = "rcdd", solver = "CrissCross")
\end{Sinput}
\begin{Soutput}
[1] TRUE
\end{Soutput}
\end{Schunk}
We have overlap in the structure vectors, there is no separation for this model specification. The coefficients and $p$~values are fine to be interpreted as they are---being of a low household rank and generally having attended many elections is associated with a very high turnout probability for the presidential primary in 2004.

\paragraph{Timings.}
This is a medium sized data set and we list the time in seconds that it takes for executing the functions above in Table~\ref{tbl:timingsb}. We compare using the default \code{"DualSimplex"} from \pkg{rcdd} with using \code{"highs"} via \pkg{ROI} as solver and backend with the following code. 
\begin{Schunk}
\begin{Sinput}
> t1.1 <- system.time(diagnose_separation(m1, backend = "rcdd",
+     solver = "DualSimplex"))
> t1.2 <- system.time(diagnose_separation(m1, backend = "ROI",
+     solver = "highs"))
\end{Sinput}
\begin{Soutput}
Warning in hi_new_solver(model): WARNING: LP matrix packed vector contains 2 |value| in [1.68952e-10, 7.82187e-10] less than or equal to 1e-09: ignored
\end{Soutput}
\begin{Sinput}
> t1.3 <- system.time(check_overlap(m2$y, model.matrix(m2), model = "bcl",
+     quick = TRUE, backend = "rcdd", solver = "DualSimplex", sequential = TRUE))
> t1.4 <- system.time(check_overlap(m2$y, model.matrix(m2), model = "bcl",
+     quick = TRUE, backend = "ROI", solver = "highs", sequential = TRUE))
\end{Sinput}
\end{Schunk}



\begin{table}
  \centering
  \begin{tabular}{llll}
    \hline
    Function & Backend & Solver & Time (seconds) \\
    \hline
    \fct{diagnose\_separation} & \pkg{rcdd} & \code{DualSimplex} & 2112.08 \\
    \fct{diagnose\_separation} & \pkg{ROI} & \code{highs} & 1996 \\
    \code{check\_overlap (quick = TRUE, sequential = TRUE)} & \pkg{rcdd} & \code{DualSimplex} & 0.42 \\
    \code{check\_overlap (quick = TRUE, sequential = TRUE)} & \pkg{ROI} & \code{highs} & 1.05\\
    \hline
  \end{tabular}
 \caption{\label{tbl:timingsb} Execution times in seconds for different functions and backend/solver setups for the Ohio voters data.}  
\end{table}

One can see that using \fct{diagnose\_seperation} takes a good while and that the choice of solver is not making much of a difference. This is due to there being only one inner function call in \fct{diagnose\_separation} that is affected by the backend/solver combination: \fct{separation\_columns}, which runs the linear program in Equation~\ref{lp2}. Hence the timings are largely similar and still largely dominated by the \pkg{rcdd} implementation. The sequential, quick check for overlap is much faster suggesting the quick check functions to be used for larger data sets. 

\subsection{Unordered categorical responses}
We adapt the alligator data of \cite{Agresti:2002} which are concerned with the primary food choices of alligators in four Florida lakes. For $n=219$ alligators we have a classification about their main food source based on their stomach content: ``Fish'', ``Invertebrates'', ``Reptiles'', ``Birds'', and ``Others''. As explanatory variables, there are the categorical variables lake of capture (``lake'') with levels ``Hancock'', ``Oklawaha'', ``Trafford'' and ``George'', the variable ``sex'' of the alligator with levels ``Male'' and ``Female'', and the variable ``size'' of the alligator with levels of ``2.3 meters or less'', and ``more than 2.3 meters''. 

We are interested in fitting a multinomial baseline-category logistic regression model to explain the food choice by the ``size'' and an interaction of ``sex'' and ``lake'' but we first check whether we actually have overlap which for this link implies that the MLE for all coefficients exists, or not. That is, we are in a pre-fit situation. We use \fct{check\_separation} with the formula interface for this.

\begin{Schunk}
\begin{Sinput}
> data("Alligators")
> check_separation(foodchoice ~ size + lake * sex, data = Alligators,
+     model = "bcl")
\end{Sinput}
\begin{Soutput}
[1] TRUE
\end{Soutput}
\end{Schunk}
So we have separation here and the MLE for some coefficients does not exist. Instead of model fitting, we therefore continue to diagnose the separation phenomenon further by checking which rows and columns are responsible.    

\begin{Schunk}
\begin{Sinput}
> scAllig <- separation_columns(foodchoice ~ size + lake * sex,
+     data = Alligators, model = "bcl")
> srAllig <- separation_rows(foodchoice ~ size + lake * sex, data = Alligators,
+     model = "bcl")
\end{Sinput}
\end{Schunk}
Separation is due to observations in these columns of the design matrix, 
\begin{Schunk}
\begin{Sinput}
> scAllig$offcols
\end{Sinput}
\begin{Soutput}
 [1] "Reptile::(Intercept)"            "Reptile::lakeOklawaha"          
 [3] "Reptile::lakeTrafford"           "Reptile::lakeGeorge"            
 [5] "Reptile::sexFemale"              "Reptile::lakeOklawaha:sexFemale"
 [7] "Reptile::lakeTrafford:sexFemale" "Reptile::lakeGeorge:sexFemale"  
 [9] "Bird::lakeOklawaha"              "Bird::lakeOklawaha:sexFemale"   
[11] "Bird::lakeGeorge:sexFemale"     
\end{Soutput}
\end{Schunk}
and the observations with these indices
\begin{Schunk}
\begin{Sinput}
> srAllig$index
\end{Sinput}
\begin{Soutput}
  [1]   1   2   3   4   5   6   7   8   9  10  11  12  13  14  15  16  17  18
 [19]  19  20  56  57  58  59  60  61  62  63  64  65  66  67  68  69  70  71
 [37]  72  73  74  75  76  77  78  79  80  81  82  83  84  85  86 157 158 159
 [55] 160 161 162 163 164 165 166 167 168 169 170 171 172 173 174 175 176 177
 [73] 178 179 180 181 182 183 184 185 186 187 188 189 190 191 192 193 194 195
 [91] 196 197 198 199 200 201 202 203 204 205 206 207 208 209 210 211 212 213
[109] 214 215 216 217 218 219
\end{Soutput}
\end{Schunk}
This is, we have separation due to several responses over all the categories, $114$ overall. If we continue with inspecting the rows in the $(X,y)$-matrix with \code{srAllig$offrows} we find a number of constellations that lead to separation phenomena. For example, in male alligators from Lake Oklawaha no birds were found and if they measured less than $2.3m$ no reptiles either. At Lake George male alligators ate no reptiles, whereas female alligators ate anything but birds and if the females were $>2.3m$ they ate neither reptiles nor birds. At Lake Trafford there was only one female $>2.3m$ and that one ate invertebrates (contributing to separation due to being a unique observation). This can then inform further data analysis steps. 

If we want to get more details, we can use \fct{diagnose\_separation} which due to the size of the data takes a while. If time is of the essence, we can also use the quick check for separation \fct{check\_separation} with argument \code{quick = TRUE}. 
\begin{Schunk}
\begin{Sinput}
> check_separation(foodchoice ~ size + lake * sex, data = Alligators,
+     model = "bcl", quick = TRUE, backend = "ROI", solver = "glpk")
\end{Sinput}
\begin{Soutput}
[1] TRUE
\end{Soutput}
\begin{Sinput}
> check_overlap(foodchoice ~ size + lake * sex, data = Alligators,
+     model = "bcl", parallel = TRUE, nc = 2, nss = 2)
\end{Sinput}
\begin{Soutput}
[1] FALSE
\end{Soutput}
\end{Schunk}



\paragraph{Timings.} We once again compare the functions used above in different setups and time the execution. We compare the default \code{"DualSimplex"} from \pkg{rcdd} with using solvers \code{"glpk"} and \code{"highs"} via \pkg{ROI} after running the following code.
\begin{Schunk}
\begin{Sinput}
> t2.1 <- system.time(check_separation(foodchoice ~ size + lake *
+     sex, data = Alligators, model = "bcl", backend = "rcdd",
+     solver = "DualSimplex"))
> t2.2 <- system.time(check_separation(foodchoice ~ size + lake *
+     sex, data = Alligators, model = "bcl", backend = "ROI", solver = "highs"))
> t2.3 <- system.time(check_separation(foodchoice ~ size + lake *
+     sex, data = Alligators, model = "bcl", backend = "ROI", solver = "glpk"))
> t2.4 <- system.time(separation_columns(foodchoice ~ size + lake *
+     sex, data = Alligators, model = "bcl", backend = "rcdd",
+     solver = "DualSimplex"))
> t2.5 <- system.time(separation_columns(foodchoice ~ size + lake *
+     sex, data = Alligators, model = "bcl", backend = "ROI", solver = "highs"))
> t2.6 <- system.time(separation_columns(foodchoice ~ size + lake *
+     sex, data = Alligators, model = "bcl", backend = "ROI", solver = "glpk"))
> t2.9 <- system.time(separation_rows(foodchoice ~ size + lake *
+     sex, data = Alligators, model = "bcl"))
> t2.7 <- system.time(check_separation(foodchoice ~ size + lake *
+     sex, data = Alligators, quick = TRUE, model = "bcl", backend = "rcdd",
+     solver = "DualSimplex"))
> t2.8 <- system.time(check_separation(foodchoice ~ size + lake *
+     sex, data = Alligators, quick = TRUE, model = "bcl", backend = "ROI",
+     solver = "glpk"))
> t2.10 <- system.time(check_separation(foodchoice ~ size + lake *
+     sex, data = Alligators, quick = TRUE, model = "bcl", backend = "ROI",
+     solver = "highs"))
\end{Sinput}
\end{Schunk}

\begin{table}
  \centering
  \begin{tabular}{llll}
    \hline
    Function & Backend & Solver & Time (seconds) \\
    \hline
    \code{check\_separation} (\code{quick = FALSE}) & \pkg{rcdd} & \code{DualSimplex} & 82.89 \\
    \code{check\_separation} (\code{quick = FALSE})& \pkg{ROI} & \code{highs} & 0.14 \\
    \code{check\_separation} (\code{quick = FALSE})& \pkg{ROI} & \code{glpk} & 0.13 \\
    \fct{separation\_columns} & \pkg{rcdd} & \code{DualSimplex} & 0.2 \\
    \fct{separation\_columns} & \pkg{ROI} & \code{highs} & 0.08 \\
    \fct{separation\_columns} & \pkg{ROI} & \code{glpk} & 0.08 \\
    \fct{separation\_rows} & \pkg{rcdd} & - & 142.91 \\
    \code{check\_separation} (\code{quick = TRUE}) & \pkg{rcdd} & \code{DualSimplex} & 0.2 \\
    \code{check\_separation} (\code{quick = TRUE}) & \pkg{ROI} & \code{highs} & 0.09 \\
    \code{check\_separation} (\code{quick = TRUE}) & \pkg{ROI} & \code{glpk} & 0.09\\
    \hline
  \end{tabular}
 \caption{\label{tbl:timingsbcl} Execution times in seconds for different functions and/or backend/solver setups for the Alligators data.}  
\end{table}

The times in seconds are given in Table~\ref{tbl:timingsbcl}. We see that \fct{check\_separation} via the dual linear program with the default setup of using \pkg{rcdd} and \code{DualSimplex} takes much longer than using alternative solvers via \pkg{ROI}. Running the \fct{separation\_columns} identification with the default takes about twice as long as with the alternative solvers. Running the separation rows identification (which can only use \pkg{rcdd}) takes 142.91 seconds. Using \fct{check\_separation} with argument \code{quick = TRUE} speeds things up considerably, especially for \pkg{rcdd} solvers.

\subsection{Ordinal responses}
\subsubsection{Cumulative link model}
\cite{drexler:2025} conducted a study on people's willingness to share health data with their social health insurance provider in Austria, with willingness measured on an ordinal scale. We use a snapshot of the original data ($n=300$) that are shipped as the data object \code{HDSS} with the \pkg{divoRce} package. This is also the data set used for illustration by \cite{Sablicaetal}.

The available variables are ``WTSSHI'', which is the respondent's willingness to share health data with the social health insurance provider as measured on an ordinal scale ranging from 1 (lowest willingness) to 4 (highest willingness), ``gender'' which is the respondents self-reported gender identification (``f'' for female, ``m'' for male and ``x'' a third category), ``chronic'' which is the respondent's self-reported number of chronic conditions (1 meaning ``one'', 2 meaning ``many'', and 3 meaning ``none''), ``trustSHI'', a Likert item capturing respondent's self-reported trust in the social health insurance provider on a scale from 1 (lowest) to 4 (highest) and ``knowledge'', a Likert item of self-reported knowledge about health data from 1 (lowest) to 4 (highest). A mosaic plot of the ordinal target variable and the trust and knowledge variables can be found in Figure~\ref{fig:wtsmosaic}.

\begin{Schunk}
\begin{Sinput}
> data("HDSS")
> library("vcd")
> vcd::mosaic(~knowledge + trustSHI + WTSSHI, highlighting = "WTSSHI",
+     direction = c("h", "v", "h"), data = HDSS, zero_size = 0)
\end{Sinput}
\end{Schunk}

\begin{figure}
\centering
\begin{Schunk}
\begin{Soutput}
Loading required package: grid
\end{Soutput}

\includegraphics[width=0.7\linewidth,height=0.35\textheight]{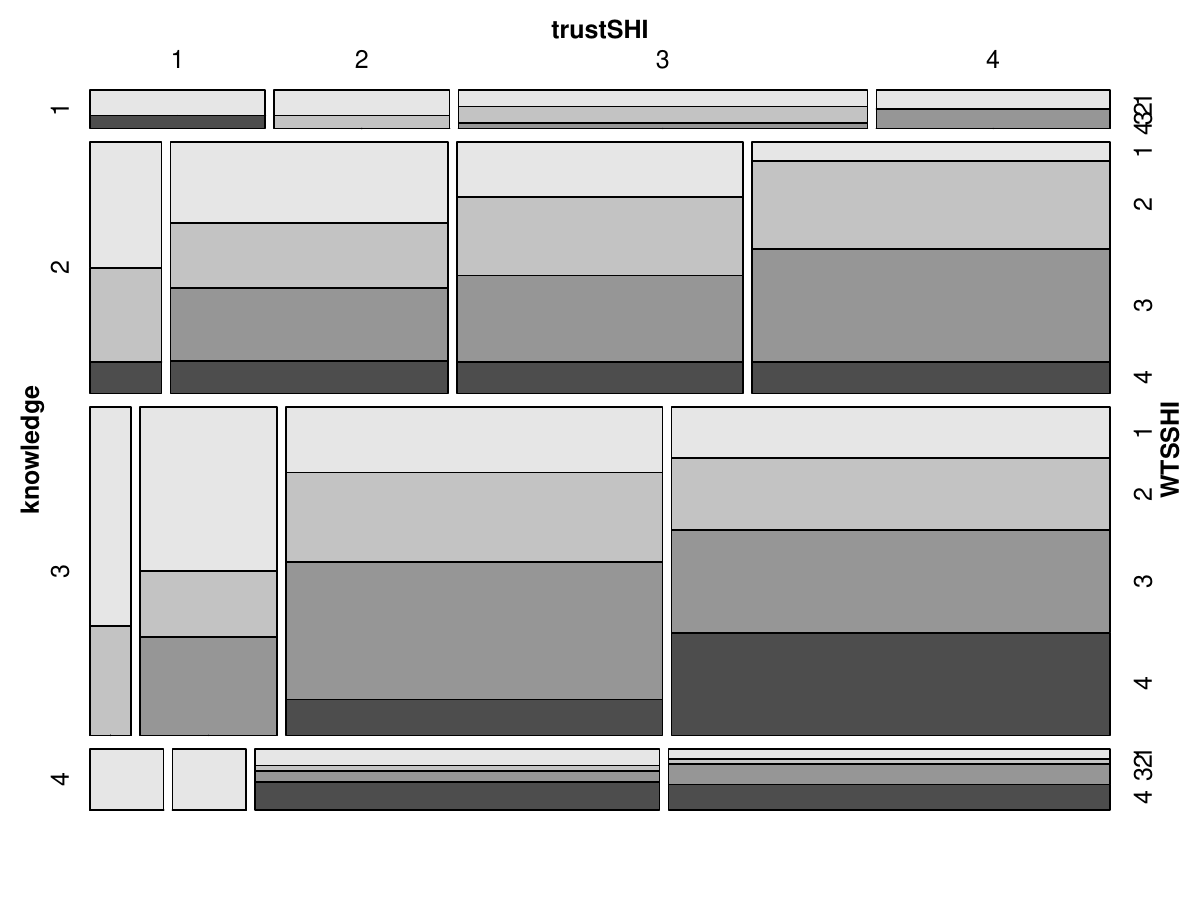} \end{Schunk}
\caption{\label{fig:wtsmosaic} A mosaic plot for the data of \cite{drexler:2025}. The mosaic is for the ordinal variable willingness to share health data (``WTSSHI'') in dependence of the categorical variables self-reported trust (``trustSHI'') and self-reported knowledge (``knowledge''). The separation phenomenon is visible in the bottom left corner.}    
\end{figure}

When analyzing the data with standard cumulative link logistic models (aka proportional odds models), one quickly runs into issues. For example, using \fct{clm} from package \pkg{ordinal} \citep{ordinalpkg} and specify a cumulative link logistic model for ``WTSSHI'' as the response with an interaction of ``trustSHI'' and ``knowledge'' in the linear predictor, we encounter numerical issues with calculating the Hessian right away, accompanied by problematic output. 
\begin{Schunk}
\begin{Sinput}
> mwtsshi <- ordinal::clm(WTSSHI ~ trustSHI * knowledge + gender +
+     chronic, data = HDSS)
\end{Sinput}
\begin{Soutput}
Warning: (1) Hessian is numerically singular: parameters are not uniquely determined 
In addition: Absolute convergence criterion was met, but relative criterion was not met
\end{Soutput}
\begin{Sinput}
> summary(mwtsshi)
\end{Sinput}
\begin{Soutput}
formula: WTSSHI ~ trustSHI * knowledge + gender + chronic
data:    HDSS

 link  threshold nobs logLik  AIC    niter max.grad cond.H 
 logit flexible  300  -372.45 788.90 18(0) 3.08e-09 4.7e+11

Coefficients:
                     Estimate Std. Error z value Pr(>|z|)
trustSHI2             -1.4603         NA      NA       NA
trustSHI3             -0.4826         NA      NA       NA
trustSHI4              0.3880         NA      NA       NA
knowledge2            -0.2531         NA      NA       NA
knowledge3            -0.9912         NA      NA       NA
knowledge4           -20.7138         NA      NA       NA
genderf               -0.7873         NA      NA       NA
genderx                0.5363         NA      NA       NA
chronic2               0.4228         NA      NA       NA
chronic3              -0.2034         NA      NA       NA
trustSHI2:knowledge2   2.3456         NA      NA       NA
trustSHI3:knowledge2   1.6121         NA      NA       NA
trustSHI4:knowledge2   1.5194         NA      NA       NA
trustSHI2:knowledge3   2.1850         NA      NA       NA
trustSHI3:knowledge3   2.5561         NA      NA       NA
trustSHI4:knowledge3   2.3265         NA      NA       NA
trustSHI2:knowledge4   0.8555         NA      NA       NA
trustSHI3:knowledge4  23.4198         NA      NA       NA
trustSHI4:knowledge4  22.8430         NA      NA       NA

Threshold coefficients:
    Estimate Std. Error z value
1|2  -0.6151         NA      NA
2|3   0.7125         NA      NA
3|4   2.4703         NA      NA
\end{Soutput}
\end{Schunk}
The culprit behind the numerical issues might be separation and we would like to check that. We can use the functions in \pkg{divoRce} to diagnose this post-fit. Note that since this is a cumulative link model, up until now it was not possible to diagnose this properly with software beyond the heuristics described in Section~\ref{sec:diag1}.  
\begin{Schunk}
\begin{Sinput}
> sd1 <- diagnose_separation(mwtsshi, backend = "rcdd", solver = "CrissCross")
> print(sd1, info = "full")
\end{Sinput}
\begin{Soutput}
Separation Check for Model: 
ordinal::clm(formula = WTSSHI ~ trustSHI * knowledge + gender + 
    chronic, data = HDSS)
Check Family: Cumulative Link Model 

Separation: TRUE 
Separation Type: Quasi-Complete Separation 

Dimension of Recession Cone: 2 
Number of Offending Columns: 4 
Number of Offending Rows: 4 
\end{Soutput}
\end{Schunk}
We find we have quasi-complete separation (the dimension of the recession cone is $2$) due to four observations (rows in the design matrix). These observations are returned as part of the object and can be accessed via
\begin{Schunk}
\begin{Sinput}
> sd1$offrows
\end{Sinput}
\end{Schunk}
(not shown). What they have in common determines the separation phenomenon: these respondents report having the highest knowledge but also report the lowest trust values (either 1 or 2 on ``trustSHI'') and are not at all willing to share their data (``WTSSHI'' of 1). This group can also be seen in the bottom left corner of Figure~\ref{fig:wtsmosaic}. This is an interesting finding as it seems it is not only the lowest trust values that comes along with a refusal to share health data, but a combination of high self-reported knowledge and low self-reported trust that gives that. 

The columns of the design matrix that are associated with separation are the partial effect for ``knowledge'' and the interaction dummies of categories 2, 3 and 4 of ``trustSHI'' with category 4 of ``knowledge'', which can be seen with \code{sd1$offcols}. For some of these columns we see highly negative coefficients in the model summary above but remarkably not for \code{trustSHI2:knowledge4}. So the separation phenomenon is only partly identifiable on the coefficient level, which clearly shows that diagnosis with the linear program is more accurate and reliable for identifying the offending columns than heuristics based on the estimated values or estimation progress. 

We can use the gathered information to sanitize the model estimation and conduct the analysis for the $296$ respondents for which we have overlap. 
\begin{Schunk}
\begin{Sinput}
> offrows <- as.numeric(row.names(sd1$offrows))
> mwtsshi2 <- ordinal::clm(WTSSHI ~ trustSHI * knowledge + gender +
+     chronic, data = HDSS[-offrows, ])
> summary(mwtsshi2)
\end{Sinput}
\begin{Soutput}
formula: WTSSHI ~ trustSHI * knowledge + gender + chronic
data:    HDSS[-offrows, ]

 link  threshold nobs logLik  AIC    niter max.grad cond.H 
 logit flexible  296  -372.45 784.90 5(0)  1.02e-12 5.2e+03

Coefficients: (2 not defined because of singularities)
                     Estimate Std. Error z value Pr(>|z|)    
trustSHI2             -1.4603     1.8473  -0.790 0.429238    
trustSHI3             -0.4826     1.5624  -0.309 0.757407    
trustSHI4              0.3880     1.6836   0.230 0.817732    
knowledge2            -0.2531     1.5576  -0.162 0.870920    
knowledge3            -0.9912     1.6382  -0.605 0.545132    
knowledge4             2.1292     1.0936   1.947 0.051546 .  
genderf               -0.7873     0.2263  -3.479 0.000503 ***
genderx                0.5363     1.6067   0.334 0.738551    
chronic2               0.4228     0.3225   1.311 0.189881    
chronic3              -0.2034     0.2612  -0.778 0.436346    
trustSHI2:knowledge2   2.3456     2.0052   1.170 0.242084    
trustSHI3:knowledge2   1.6121     1.7334   0.930 0.352339    
trustSHI4:knowledge2   1.5194     1.8419   0.825 0.409449    
trustSHI2:knowledge3   2.1850     2.0746   1.053 0.292244    
trustSHI3:knowledge3   2.5561     1.7947   1.424 0.154368    
trustSHI4:knowledge3   2.3265     1.9018   1.223 0.221214    
trustSHI2:knowledge4       NA         NA      NA       NA    
trustSHI3:knowledge4   0.5768     1.4465   0.399 0.690052    
trustSHI4:knowledge4       NA         NA      NA       NA    
---
Signif. codes:  0 '***' 0.001 '**' 0.01 '*' 0.05 '.' 0.1 ' ' 1

Threshold coefficients:
    Estimate Std. Error z value
1|2  -0.6151     1.4314  -0.430
2|3   0.7125     1.4357   0.496
3|4   2.4703     1.4399   1.716
\end{Soutput}
\end{Schunk}
This model has overlap 
\begin{Schunk}
\begin{Sinput}
> check_overlap(mwtsshi2, parallel = TRUE, nc = 2, nss = 2)
\end{Sinput}
\begin{Soutput}
[1] TRUE
\end{Soutput}
\end{Schunk}
We now see that, except for the lowest knowledge category, higher knowledge is generally associated with a higher willingness to share the data and that that association is higher the more trust the respondent reports, quite the opposite to the pattern we observe for the observations that are associated with separation. Note, we while we have no separation anymore, the design matrix is now singular. This also illustrates that the two situations of nonexisting and nonidentifiable coefficients need to be distinguished.

As a side note, for these data all other ordinal model families as well as the baseline-category model would also show separation for the same linear predictor. We can confirm this in a pre-fit situation for example by supplying the response vector and the design matrix (extracted from the model object) to \fct{diagnose\_separation} for different models. For example,
\begin{Schunk}
\begin{Sinput}
> Xwts <- model.matrix(mwtsshi)$X
> ywts <- HDSS$WTSSHI
> check_separation(ywts, Xwts, model = "os")
> check_separation(ywts, XRwts, model = "acl")
> check_separation(ywts, Xwts, model = "sl")
> check_separation(unclass(ywts), Xwts, model = "bcl")
\end{Sinput}
\end{Schunk}
\paragraph{Timings.}
Table~\ref{tbl:timingscl} compares running the diagnostics utlizing \code{"CrissCross"} from \pkg{rcdd} with using \code{"highs"} and \code{"lpsolve"} via \pkg{ROI} as solver and backend. Results are obtained via this code. 
\begin{Schunk}
\begin{Sinput}
> t3.1 <- system.time(diagnose_separation(mwtsshi, backend = "rcdd",
+     solver = "CrissCross"))
> t3.2 <- system.time(diagnose_separation(mwtsshi, backend = "ROI",
+     solver = "highs"))
> t3.3 <- system.time(diagnose_separation(mwtsshi, backend = "ROI",
+     solver = "lpsolve"))
> t4.1 <- system.time(check_overlap(mwtsshi2, parallel = FALSE,
+     backend = "rcdd", solver = "CrissCross"))
> t4.2 <- system.time(check_overlap(mwtsshi2, parallel = TRUE,
+     nc = 2, nss = 2, backend = "rcdd", solver = "CrissCross"))
> t4.3 <- system.time(check_overlap(mwtsshi2, parallel = TRUE,
+     nc = 2, nss = 2, backend = "ROI", solver = "highs"))
> t4.4 <- system.time(check_overlap(mwtsshi2, parallel = TRUE,
+     nc = 2, nss = 2, backend = "ROI", solver = "lpsolve"))
\end{Sinput}
\end{Schunk}

\begin{table}
  \centering
  \begin{tabular}{llll}
    \hline
    Function & Backend & Solver & Time (seconds) \\
    \hline
    \fct{diagnose\_separation} & \pkg{rcdd} & \code{CrissCross} & 13.13 \\
    \fct{diagnose\_separation}& \pkg{ROI} & \code{highs} & 13.31 \\
    \fct{diagnose\_separation}& \pkg{ROI} & \code{lpsolve} & 14.88 \\
    \code{check\_overlap()} & \pkg{rcdd} & \code{CrissCross} & 4.42 \\
    \code{check\_overlap(parallel = TRUE)}& \pkg{rcdd} & \code{CrissCross} & 5.13 \\
    \code{check\_overlap(parallel = TRUE)}& \pkg{ROI} & \code{highs} & 0.13 \\
     \code{check\_overlap(parallel = TRUE)}& \pkg{ROI} & \code{lpsolve} & 0.22 \\
    \hline
  \end{tabular}
 \caption{\label{tbl:timingscl} Execution times in seconds for different backend/solver setups for diagnosing the HDSS data.}  
\end{table}
The timings show the same pattern as before. Of note is that the parallel execution takes about as long as the non-parallel one which might be due to the first subset already showing overlap.  

More examples (including for adjacent-category link models, ordered stereotype models and sequential models) can be found in the package vignette.

\section{Discussion}
This article discussed computational, numerical and practical aspects of separation phenomena in categorical response models when maximum likelihood estimation is employed. We covered categorical response models for binary and polytomous nominal and ordinal responses. We discussed consequences of separation and how the problems induced by separation can be addressed. We further laid out how the most popular commercial statistical software packages, \proglang{SAS}, \proglang{SPSS}, \proglang{Stata} handle separation and what is currently available in \proglang{R}.

Furthermore we introduced our own \proglang{R} package \pkg{divoRce} that turns the structure vector theory of \cite{Sablicaetal} into usable diagnostic tools. The distinguishing features of \pkg{divoRce} are that the software works for almost any categorical response models family in existence and almost any type of link construction, link type and underlying error distributions. Our software allows to check for separation and detect the variables or design matrix columns responsible with accurate linear programs that can be solved with the solvers in \pkg{rcdd} or any of the solvers available for \pkg{ROI}. It further enables a user to diagnose the type of separation (complete separation, quasi-complete separation or no separation aka overlap), to characterize the recession cone and its dimension and to identify the observations that contribute to separation with methods from computational geometry. All of this can be done in floating-point arithmetic or in rational arithmetic with arbitrary precision. The package \pkg{divoRce} is created in a modular fashion and offers a computational infrastructure that can easily be extended by third parties by providing wrappers and methods for their own implementations of categorical response models.  

We further illustrated the usage of \pkg{divoRce} functionality with three examples of categorical response model families and also explored the time it took to do the calculations, providing a manual for the usage of the package. 

The work in this article provides software access to the unified theoretical framework of \cite{Sablicaetal} and allows users, data analysts and software developers to quickly, reliably and accurately diagnose problems of possible nonexistence of the MLE in their models due to separation, enabling them to routinely check for separation as part of a model diagnostic and estimation validation workflow whenever categorical response models are employed. Similar to checking goodness-of-fit or doing residual diagnostics, we recommend this as a standard statistical practice going forward because the separation that burns our hearts is the root of all our disease\footnote{From ``We are torn wide open'' by Neurosis (2026).}. 
\bibliography{../polybib}

\begin{thebibliography}{44}
\newcommand{\enquote}[1]{``#1''}
\providecommand{\natexlab}[1]{#1}
\providecommand{\url}[1]{\texttt{#1}}
\providecommand{\urlprefix}{URL }
\expandafter\ifx\csname urlstyle\endcsname\relax
  \providecommand{\doi}[1]{doi:\discretionary{}{}{}#1}\else
  \providecommand{\doi}{doi:\discretionary{}{}{}\begingroup
  \urlstyle{rm}\Url}\fi
\providecommand{\eprint}[2][]{\url{#2}}

\bibitem[{Agresti(2022)}]{Agresti:2002}
Agresti A (2022).
\newblock \emph{Categorical Data Analysis}.
\newblock 2nd edition. John Wiley \& Sons, Hoboken, New Jersey.
\newblock \doi{10.1007/978-3-642-04898-2_161}.

\bibitem[{Albert and Anderson(1984)}]{polytomous:Albert+Anderson:1984}
Albert A, Anderson JA (1984).
\newblock \enquote{On the Existence of Maximum Likelihood Estimates in Logistic
  Regression Models.}
\newblock \emph{Biometrika}, \textbf{71}(1), 1--10.
\newblock \doi{10.1093/biomet/71.1.1}.

\bibitem[{Allison(2008)}]{allison}
Allison P (2008).
\newblock \enquote{Convergence Failures in Logistic Regression.}
\newblock SAS Global Forum 2008.
\newblock
  \urlprefix\url{https://support.sas.com/resources/papers/proceedings/pdfs/sgf2008/360-2008.pdf}.

\bibitem[{Berkelaar \emph{et~al.}(2004)Berkelaar, Eikland, and
  Notebaert}]{lpsolve}
Berkelaar M, Eikland K, Notebaert P (2004).
\newblock \enquote{\pkg{lp\_solve} 5.5, Open Source (Mixed-Integer) Linear
  Programming System.}
\newblock \urlprefix\url{http://lpsolve.sourceforge.net/5.5/}.

\bibitem[{Christensen(2023)}]{ordinalpkg}
Christensen RHB (2023).
\newblock \emph{\pkg{ordinal}---Regression Models for Ordinal Data}.
\newblock \doi{10.32614/CRAN.package.ordinal}.
\newblock \proglang{R} package version 2023.12-4.1,
  \urlprefix\url{https://CRAN.R-project.org/package=ordinal}.

\bibitem[{Coveney(2008)}]{firthlogit}
Coveney J (2008).
\newblock \enquote{\pkg{FIRTHLOGIT}: \proglang{Stata} Module to Calculate Bias
  Reduction in Logistic Regression.}
\newblock Statistical Software Components, Boston College Department of
  Economics.
\newblock \urlprefix\url{https://ideas.repec.org/c/boc/bocode/s456948.html}.

\bibitem[{Drexler(2026)}]{drexler:2025}
Drexler S (2026).
\newblock \emph{Patient-reported and Register-based Indicators for System-wide
  Primary Care Quality Evaluation: An Analysis of Applicability, Benefits,
  Limitations, and Mitigation Strategies}.
\newblock Phd thesis, WU Vienna University of Economics and Business, Vienna,
  Austria.

\bibitem[{Fienberg and Rinaldo(2012)}]{Fienberg+Rinaldo:2012}
Fienberg SE, Rinaldo A (2012).
\newblock \enquote{Maximum Likelihood Estimation in Log-linear Models.}
\newblock \emph{The Annals of Statistics}, \textbf{40}(2), 996--1023.
\newblock \doi{10.1214/12-AOS986}.

\bibitem[{Firth(1993)}]{firth1993bias}
Firth D (1993).
\newblock \enquote{Bias Reduction of Maximum Likelihood Estimates.}
\newblock \emph{Biometrika}, \textbf{80}(1), 27--38.

\bibitem[{{Free Software Foundation}(2023)}]{GMP}
{Free Software Foundation} (2023).
\newblock \enquote{{GNU} Multiple Precision Library.}
\newblock \urlprefix\url{https://gmplib.org/}.

\bibitem[{Fukuda(2025)}]{cddlib}
Fukuda K (2025).
\newblock \enquote{\pkg{cddlib}: Double Description Method for Polyhedral
  Representation Conversion.}
\newblock \urlprefix\url{https://github.com/cddlib/cddlib}.

\bibitem[{Gelman \emph{et~al.}(2008)Gelman, Jakulin, Pittau, and
  Su}]{Gelman2008}
Gelman A, Jakulin A, Pittau MG, Su YS (2008).
\newblock \enquote{A Weakly Informative Default Prior Distribution for Logistic
  and Other Regression Models.}
\newblock \emph{The Annals of Applied Statistics}, \textbf{2}(4), 1360--1383.
\newblock \doi{10.1214/08-AOAS191}.

\bibitem[{Geyer(2009)}]{geyer2009likelihood}
Geyer CJ (2009).
\newblock \enquote{Likelihood Inference in Exponential Families and Directions
  of Recession.}
\newblock \emph{Electronic Journal of Statistics}, \textbf{3}, 259--289.
\newblock \doi{10.1214/08-EJS349}.

\bibitem[{Geyer(2021)}]{geyervinny}
Geyer CJ (2021).
\newblock \emph{Using the \pkg{rcdd} Package}.
\newblock \proglang{R} Package Vignette,
  \urlprefix\url{https://cran.r-project.org/web/packages/rcdd/vignettes/vinny.pdf}.

\bibitem[{Geyer \emph{et~al.}(2026)Geyer, Meeden, and Fukuda}]{rcdd}
Geyer CJ, Meeden GD, Fukuda K (2026).
\newblock \emph{\pkg{rcdd}: Computational Geometry}.
\newblock \doi{10.32614/CRAN.package.rcdd}.
\newblock \proglang{R} package version 1.6-1,
  \urlprefix\url{https://CRAN.R-project.org/package=rcdd}.

\bibitem[{Greenland and Mansournia(2015)}]{Greenland2015}
Greenland S, Mansournia MA (2015).
\newblock \enquote{Penalization, Bias Reduction, and Default Priors in Logistic
  and Related Categorical and Survival Regressions.}
\newblock \emph{Statistics in Medicine}, \textbf{34}(23), 3133--3143.
\newblock \doi{10.1002/sim.6537}.

\bibitem[{Hauck and Donner(1977)}]{hauckdonner}
Hauck WW, Donner A (1977).
\newblock \enquote{Wald's Test as Applied to Hypotheses in Logit Analysis.}
\newblock \emph{Journal of the American Statistical Association},
  \textbf{72}(360a), 851--853.
\newblock \doi{10.1080/01621459.1977.10479969}.

\bibitem[{Heinze \emph{et~al.}(2025)Heinze, Ploner, Jiricka, and
  Steiner}]{logistf}
Heinze G, Ploner M, Jiricka L, Steiner G (2025).
\newblock \emph{\pkg{logistf}: Firth's Bias-Reduced Logistic Regression}.
\newblock \doi{10.32614/CRAN.package.logistf}.
\newblock \proglang{R} package version 1.26.1,
  \urlprefix\url{https://CRAN.R-project.org/package=logistf}.

\bibitem[{Heinze and Schemper(2002)}]{heinze2002solution}
Heinze G, Schemper M (2002).
\newblock \enquote{A Solution to the Problem of Separation in Logistic
  Regression.}
\newblock \emph{Statistics in Medicine}, \textbf{21}(16), 2409--2419.
\newblock \doi{10.1002/sim.1047}.

\bibitem[{Huangfu and Hall(2015)}]{highs}
Huangfu Q, Hall JAJ (2015).
\newblock \enquote{Parallelizing the Dual Revised Simplex Method.}
\newblock \urlprefix\url{https://arxiv.org/abs/1503.01889}.

\bibitem[{{IBM Corp.}(2016)}]{spss24}
{IBM Corp} (2016).
\newblock \enquote{GENLIN Algorithms.}
\newblock In \emph{IBM SPSS Statistics for Windows, Version 24.0}. IBM Corp.,
  Armonk, NY.
\newblock
  \urlprefix\url{https://public.dhe.ibm.com/software/analytics/spss/documentation/statistics/24.0/en/client/Manuals/IBM_SPSS_Statistics_Algorithms.pdf}.

\bibitem[{Kaufmann(1988)}]{polytomous:Kaufmann:1988}
Kaufmann H (1988).
\newblock \enquote{On Existence and Uniqueness of Maximum Likelihood Estimates
  in Quantal and Ordinal Response Models.}
\newblock \emph{Metrika}, \textbf{35}, 291--313.
\newblock \doi{10.1007/BF02613318}.

\bibitem[{Kleiber and Zeileis(2008)}]{AER}
Kleiber C, Zeileis A (2008).
\newblock \emph{Applied Econometrics with \proglang{R}}.
\newblock Springer-Verlag, New York.
\newblock \doi{10.1007/978-0-387-77318-6}.

\bibitem[{Konis(2007)}]{konis2007}
Konis K (2007).
\newblock \emph{Linear Programming Algorithms for Detecting Separated Data in
  Binary Logistic Regression Models}.
\newblock Phd thesis, University of Oxford.
\newblock \doi{10.5287/ora-yrojgqb76}.

\bibitem[{Kosmidis(2023{\natexlab{a}})}]{kosmidisACL}
Kosmidis I (2023{\natexlab{a}}).
\newblock \enquote{Mean and Median Bias Reduction: A Concise Review and
  Application to Adjacent-Categories Logit Models.}
\newblock In M~Kateri, I~Moustaki (eds.), \emph{Trends and Challenges in
  Categorical Data Analysis: Statistical Modelling and Interpretation}, pp.
  177--197. Springer-Verlag, Cham.
\newblock \doi{10.1007/978-3-031-31186-4_6}.

\bibitem[{Kosmidis(2023{\natexlab{b}})}]{brglm2}
Kosmidis I (2023{\natexlab{b}}).
\newblock \emph{\pkg{brglm2}: Bias Reduction in Generalized Linear Models}.
\newblock \doi{10.32614/CRAN.package.brglm2}.
\newblock \proglang{R} package version 0.9.2,
  \urlprefix\url{https://CRAN.R-project.org/package=brglm2}.

\bibitem[{Kosmidis and Firth(2021)}]{kosmidis2021}
Kosmidis I, Firth D (2021).
\newblock \enquote{Jeffreys-prior Penalty, Finiteness and Shrinkage in
  Binomial-response Generalized Linear Models.}
\newblock \emph{Biometrika}, \textbf{108}, 71--82.
\newblock \doi{10.1093/biomet/asaa052}.

\bibitem[{Kosmidis \emph{et~al.}(2026)Kosmidis, Schumacher, and
  Schwendinger}]{detectseparation}
Kosmidis I, Schumacher D, Schwendinger F (2026).
\newblock \emph{\pkg{detectseparation}: Detect and Check for Separation and
  Infinite Maximum Likelihood Estimates}.
\newblock \doi{10.32614/CRAN.package.detectseparation}.
\newblock \proglang{R} package version 0.4.0,
  \urlprefix\url{https://CRAN.R-project.org/package=detectseparation}.

\bibitem[{Makhorin(2020)}]{glpk}
Makhorin A (2020).
\newblock \enquote{GLPK (GNU Linear Programming Kit).}
\newblock \urlprefix\url{http://www.gnu.org/software/glpk/glpk.html}.

\bibitem[{Rainey(2016)}]{Rainey2016}
Rainey C (2016).
\newblock \enquote{Dealing with Separation in Logistic Regression Models.}
\newblock \emph{Political Analysis}, \textbf{24}, 339--355.
\newblock \doi{10.1093/pan/mpw014}.

\bibitem[{Rusch \emph{et~al.}(2013)Rusch, Lee, Hornik, Jank, and
  Zeileis}]{Ruschetal:2013}
Rusch T, Lee I, Hornik K, Jank W, Zeileis A (2013).
\newblock \enquote{Influencing Elections with Statistics: Targeting Voters with
  Logistic Regression Trees.}
\newblock \emph{The Annals of Applied Statistics}, pp. 1612--1639.
\newblock \doi{10.1214/13-AOAS648}.

\bibitem[{Rusch \emph{et~al.}(2026)Rusch, Sablica, and
  Hornik}]{separationcheck}
Rusch T, Sablica L, Hornik K (2026).
\newblock \emph{\pkg{divoRce}: Diagnosing Separation Phenomena in Latent
  Polyhedral Categorical Response Models}.
\newblock \proglang{R} package version 1.0-2,
  \urlprefix\url{https://cran.r-project.org/package=divoRce}.

\bibitem[{Sablica \emph{et~al.}(2026)Sablica, Hornik, and Rusch}]{Sablicaetal}
Sablica L, Hornik K, Rusch T (2026).
\newblock \enquote{The Geometry of Existence and Uniqueness of Maximum
  Likelihood Estimation in Categorical Response Models.}
\newblock \eprint{2610.08100},
  \urlprefix\url{https://arxiv.org/abs/2610.08100}.

\bibitem[{{SAS Institute Inc.}(2023)}]{sas-sep}
{SAS Institute Inc} (2023).
\newblock \enquote{{The LOGISTIC Procedure}.}
\newblock In \emph{SAS/STAT 15.3 User's Guide}. {SAS Institute Inc.}, Cary, NC.
\newblock
  \urlprefix\url{https://documentation.sas.com/doc/en/statug/15.3/statug_logistic_toc.htm}.

\bibitem[{Schwendinger(2023)}]{lpsolveROI}
Schwendinger F (2023).
\newblock \emph{\pkg{ROI.plugin.lpsolve}{: 'lp\_solve' Plugin for the
  \proglang{R} Optimization Infrastructure}}.
\newblock \doi{10.32614/CRAN.package.ROI.plugin.lpsolve}.
\newblock \proglang{R} package version 1.0-2.

\bibitem[{Schwendinger(2025)}]{highsROI}
Schwendinger F (2025).
\newblock \emph{\pkg{ROI.plugin.highs}{: 'HiGHS' Plugin for the \proglang{R}
  Optimization Infrastructure}}.
\newblock \doi{10.32614/CRAN.package.ROI.plugin.highs}.
\newblock \proglang{R} package version 1.0-4.

\bibitem[{Schwendinger \emph{et~al.}(2021)Schwendinger, Gr\"un, and
  Hornik}]{schwendlogbinom}
Schwendinger F, Gr\"un B, Hornik K (2021).
\newblock \enquote{A Comparison of Optimization Solvers for Log Binomial
  Regression Including Conic Programming.}
\newblock \emph{Computational Statistics}, \textbf{36}, 1721--1754.
\newblock \doi{10.1007/s00180-021-01084-5}.

\bibitem[{Silvapulle(1981)}]{polytomous:Silvapulle:1981}
Silvapulle MJ (1981).
\newblock \enquote{On the Existence of Maximum Likelihood Estimators for the
  Binomial Response Model.}
\newblock \emph{Journal of the Royal Statistical Society, Series B (Statistical
  Methodology)}, \textbf{43}(3), 310--313.
\newblock \doi{10.1111/j.2517-6161.1981.tb01676.x}.

\bibitem[{Speckman \emph{et~al.}(2009)Speckman, Lee, and
  Sun}]{polytomous:Speckman+Lee+Sun:2009}
Speckman PL, Lee J, Sun D (2009).
\newblock \enquote{Existence of the {MLE} and Propriety of Posteriors for a
  General Multinomial Choice Model.}
\newblock \emph{Statistica Sinica}, \textbf{19}, 731--748.
\newblock \urlprefix\url{https://www.jstor.org/stable/24308853}.

\bibitem[{StataCorp(2025)}]{rlogit}
StataCorp (2025).
\newblock \emph{\pkg{logit} - Logistic Regression, Reporting Coefficients}.
\newblock College Station: TX.
\newblock \urlprefix\url{https://www.stata.com/manuals/rlogit.pdf}.

\bibitem[{Theussl(2020)}]{glpkROI}
Theussl S (2020).
\newblock \emph{\pkg{ROI.plugin.glpk}{: 'ROI' Plug-in 'GLPK'}}.
\newblock \doi{10.32614/CRAN.package.ROI.plugin.glpk}.
\newblock \proglang{R} package version 1.0-0.

\bibitem[{Theussl \emph{et~al.}(2020)Theussl, Schwendinger, and
  Hornik}]{ROIpap}
Theussl S, Schwendinger F, Hornik K (2020).
\newblock \enquote{\pkg{ROI}: An Extensible \proglang{R} Optimization
  Infrastructure.}
\newblock \emph{Journal of Statistical Software}, \textbf{94}(15), 1--64.
\newblock \doi{10.18637/jss.v094.i15}.

\bibitem[{Zamar \emph{et~al.}(2007)Zamar, McNeney, and Graham}]{elrm}
Zamar D, McNeney B, Graham J (2007).
\newblock \enquote{\pkg{elrm}: Software Implementing Exact-like Inference for
  Logistic Regression Models.}
\newblock \emph{Journal of Statistical Software}, \textbf{21}(3).
\newblock \doi{10.18637/jss.v021.i03}.

\bibitem[{Zou and Hastie(2005)}]{zou2005regularization}
Zou H, Hastie T (2005).
\newblock \enquote{Regularization and Variable Selection via the Elastic Net.}
\newblock \emph{Journal of the Royal Statistical Society, Series B (Statistical
  Methodology)}, \textbf{67}(2), 301--320.
\newblock \doi{10.1111/j.1467-9868.2005.00503.x}.

\end{thebibliography}

\end{document}